\documentclass[pdflatex,sn-nature]{sn-jnl}

\usepackage{graphicx}
\usepackage{multirow}
\usepackage{amsmath,amssymb,amsfonts}
\usepackage{amsthm}
\usepackage{mathrsfs}
\usepackage[title]{appendix}
\usepackage{xcolor}
\usepackage{textcomp}
\usepackage{manyfoot}
\usepackage{booktabs}
\usepackage{algorithm}
\usepackage{algorithmicx}
\usepackage{algpseudocode}
\usepackage{listings}
\usepackage{subcaption}
\usepackage[normalem]{ulem}

\renewcommand{\vec}[1]{\boldsymbol{#1}}
\renewcommand{\Vec}[1]{\boldsymbol{#1}}

\begin{document}

\title[Article Title]{In-plane ferromagnetism-driven topological nodal-point superconductivity with tilted Weyl cones}

\author[1,2]{\fnm{Maciej} \sur{Bazarnik}}
\author*[3,4]{\fnm{Levente} \sur{R\'{o}zsa}}\email{rozsa.levente@wigner.hun-ren.hu}
\author[5,6]{\fnm{Ioannis} \sur{Ioannidis}}
\author[7]{\fnm{Eric} \sur{Mascot}}
\author[1]{\fnm{Philip} \sur{Beck}}
\author[3,4,8]{\fnm{Kriszti\'{a}n} \sur{Palot\'{a}s}}
\author[4]{\fnm{Andr\'{a}s} \sur{De\'{a}k}}
\author[4,9]{\fnm{L\'{a}szl\'{o}} \sur{Szunyogh}}
\author[7]{\fnm{Stephan} \sur{Rachel}}
\author[5,6]{\fnm{Thore} \sur{Posske}}
\author[1]{\fnm{Roland} \sur{Wiesendanger}}
\author[1,6]{\fnm{Jens} \sur{Wiebe}}
\author[1,6]{\fnm{Kirsten} \sur{von Bergmann}}
\author*[1,10]{\fnm{Roberto} \sur{Lo Conte}}\email{r.lo.conte@rug.nl}

\affil[1]{\orgdiv{Institute of Nanostructure and Solid State Physics}, \orgname{University of Hamburg}, \postcode{20355} \city{Hamburg}, \country{Germany}}
\affil[2]{\orgdiv{Institute of Physics}, \orgname{University of M\"unster}, \postcode{48149} \city{M\"unster}, \country{Germany}}
\affil[3]{\orgdiv{Department of Theoretical Solid State Physics}, \orgname{HUN-REN Wigner Research Centre for Physics}, \postcode{H-1525} \city{Budapest}, \country{Hungary}}
\affil[4]{\orgdiv{Department of Theoretical Physics}, \orgname{Budapest University of Technology and Economics}, \street{M\H{u}egyetem rakpart 3.}, \postcode{H-1111} \city{Budapest}, \country{Hungary}}
\affil[5]{\orgdiv{
I. Institute for Theoretical Physics}, \orgname{University of Hamburg}, \postcode{D-22607} \city{Hamburg}, \country{Germany}}
\affil[6]{\orgdiv{
The Hamburg Centre for Ultrafast Imaging, Luruper Chaussee 149}, \orgname{University of Hamburg}, \postcode{D-22761} \city{Hamburg}, \country{Germany}}
\affil[7]{\orgdiv{School of Physics}, \orgname{University of Melbourne}, \postcode{3010} \city{Victoria}, \country{Australia}}
\affil[8]{\orgdiv{HUN-REN-SZTE Reaction Kinetics and Surface Chemistry Research Group}, \orgname{University of Szeged}, \postcode{H-6720} \city{Szeged}, \country{Hungary}}
\affil[9]{\orgdiv{HUN-REN-BME Condensed Matter Research Group}, \orgname{Budapest University of Technology and Economics}, \street{M\H{u}egyetem rakpart 3.}, \postcode{H-1111} \city{Budapest}, \country{Hungary}}
\affil[10]{\orgdiv{Zernike Institute for Advanced Materials}, \orgname{University of Groningen}, \postcode{9747AG} \city{Groningen}, \country{The Netherlands}}


\abstract{The potential application of topological superconductivity in quantum transport and quantum information has fueled an intense investigation of hybrid materials 
with emergent electronic properties, 
including magnet-superconductor heterostructures. Here, we report evidence of a topological nodal-point superconducting phase in a one-atom-thick in-plane ferromagnet in direct proximity to a conventional $s$-wave superconductor. Low-temperature scanning tunneling spectroscopy data reveal the presence of a double-peak low-energy feature in the local density of states of the hybrid system, which is rationalized via 
model calculations to be an emergent topological nodal-point superconducting phase with tilted Weyl cones. Our results further establish the combination of in-plane ferromagnetism and conventional superconductivity as a route to design two-dimensional topological quantum phases.}


\maketitle

\section{Introduction}\label{intro}

The combination of magnetism and superconductivity is a very promising approach for the design of new topological materials applicable in future quantum technologies.
In particular, magnet-superconductor heterostructures (MSHs) are expected to be an ideal platform for the stabilization of topological superconductivity~\cite{Li2016,LoConteRivNuovCim2024,SteffensenPRR2022}, whose quantized and especially localized electronic and thermal transport properties and non-Abelian excitations are a key feature in next generation quantum technology applications~\cite{NayakREV-MOD-PHYS2008,sato2017TopologicalSuperconductorsReview,AliceaNAT-PHYS2011,SauPRL2010}.
The emergence of two-dimensional (2D) topological superconductivity in MSHs has been explored extensively in the recent years~\cite{MenardNAT-COMM2017,Palacio-MoralesSCI-ADV2019,KezilebiekeNAT2020,LoContePRB2022,Bazarnik2023-natcomm,SoldiniNAT-PHYS2023,bruning_ACS2025,Zahner2025}.
It has been shown that the combination of 2D ferromagnets with out-of-plane spin alignment and $s$-wave superconductors gives rise to a gapped topological phase with chiral edge modes~\cite{Li2016,Palacio-MoralesSCI-ADV2019,KezilebiekeNAT2020,MenardNAT-COMM2017}.
Such a topological phase is also expected to arise in some non-coplanar spin textures proximitized to superconductors~\cite{BedowPRB2020,MascotQ-MAT2021,nickel2025-2}.
More recently, the attention has shifted towards the exploration of MSHs hosting 2D antiferromagnetic ground states with out-of-plane spin alignment~\cite{LoContePRB2022,Bazarnik2023-natcomm,SoldiniNAT-PHYS2023,KieuPRB2023,Zahner2025}, where the emergence of a gapless topological phase was experimentally observed.
The characteristics of these observed gapless superconducting states are the presence of pairs of nodal points in the Brillouin zone energetically located at the Fermi level, accompanied by the emergence of edge modes located only at edges along specific crystallographic directions.
Nodal-point topological superconductivity has also been predicted~\cite{NakosaiPRB2013} and recently experimentally observed in a MSH hosting a coplanar spin spiral ground state~\cite{bruning_ACS2025}.

Gapless nodal-point superconducting phases in 2D MSHs with antiferromagnetic and non-collinear spin textures are understood to emerge due to time-reversal symmetry. 
While the magnetic structure breaks physical time-reversal symmetry, its combination with a mirroring or a 180$^{\circ}$ rotation restores an effective time-reversal symmetry~\cite{Heimes2015,SoldiniNAT-PHYS2023}. 
Together with the particle-hole constraint of superconductors, this places these types of systems in the BDI symmetry class~\cite{Chiu2016} where topologically protected nodal points form in two dimensions at the meeting point of Weyl cones~\cite{Beri2010}.
In addition, nodal points with opposite winding numbers are connected via flat surface modes, which can be experimentally observed at the one-dimensional edges of the 2D magnetic islands.

So far, 2D MSHs with in-plane ferromagnetic order have been explored less experimentally.
Theoretical reports proposed in-plane ferromagnet-superconductor heterostructures as an alternative platform for the engineering of nodal point topological superconducting phases~\cite{Godzik2020}, highlighting van der Waals (vdW) heterostructures as the ideal platform for the exploration of this emergent quantum phase.
In-plane ferromagnets with a twofold rotational crystal symmetry around the out-of-plane direction also possess an effective time-reversal symmetry as a combination of this rotation and physical time reversal.
Hence, they as well belong to the BDI symmetry class where nodal points are expected to appear. 
A recent experimental study of in-plane ferromagnetic CrCl$_3$ on superconducting NbSe$_2$ showed the emergence of Yu-Shiba-Rusinov (YSR) bound states~\cite{YuACT-PHYS-SIN1965,ShibaPROG-THEO-PHYS1968,Rusinov1969} at the edges of the in-plane ferromagnetic CrCl$_3$ island~\cite{Cuperus2025}.
However, no evidence of emergent topological properties was obtained.

Here, we report on experimental evidence for a topological nodal-point superconducting phase in an in-plane ferromagnet-superconductor heterostructure. 
The system consists of Co monolayer islands on a superconducting Nb(110) single crystal substrate.
Low-temperature scanning tunneling microscopy (STM) measurements reveal the formation of a buckled Co monolayer on the Nb(110) surface, which results in a denser Co layer compared to the bcc(110) surface of the substrate.
First-principles density-functional theory (DFT) calculations confirm a reconstructed $(3\times1)$ Co monolayer on the Nb(110) surface as the energetically most favorable structural configuration, which is found to host an in-plane ferromagnetic ground state with magnetic easy axis along the buckling direction, based on atomistic spin-dynamics simulations.
Low-temperature scanning tunneling spectroscopy (STS) measurements reveal the presence of a double-peak feature around the Fermi level in the differential tunneling conductance (d\textit{I}/d\textit{V}) of the Co/Nb(110) sample, as well as an enhanced low-energy d\textit{I}/d\textit{V} at the edges of the Co monolayer islands.
An analytical low-energy model and tight-binding calculations allow us to interpret the observed spectroscopic features as evidence of a nodal-point superconducting phase where the in-plane magnetization moves the nodal points away from the Fermi level, resulting in the tilting of the Weyl cones of the emergent electronic band structure. 
While for the present system type-I Weyl nodes are predicted, the analytical low-energy model  also unveils in-plane ferromagnet-superconductor hybrids as a promising platform for the realization of topological nodal-point superconductors harboring type-II Weyl physics~\cite{SoluyanovNAT2015}.

\section{Results}\label{results}
\subsection{Structural and magnetic properties}\label{struANDmag}
\begin{figure}[ht!]
\centering
\includegraphics[width=1\linewidth]{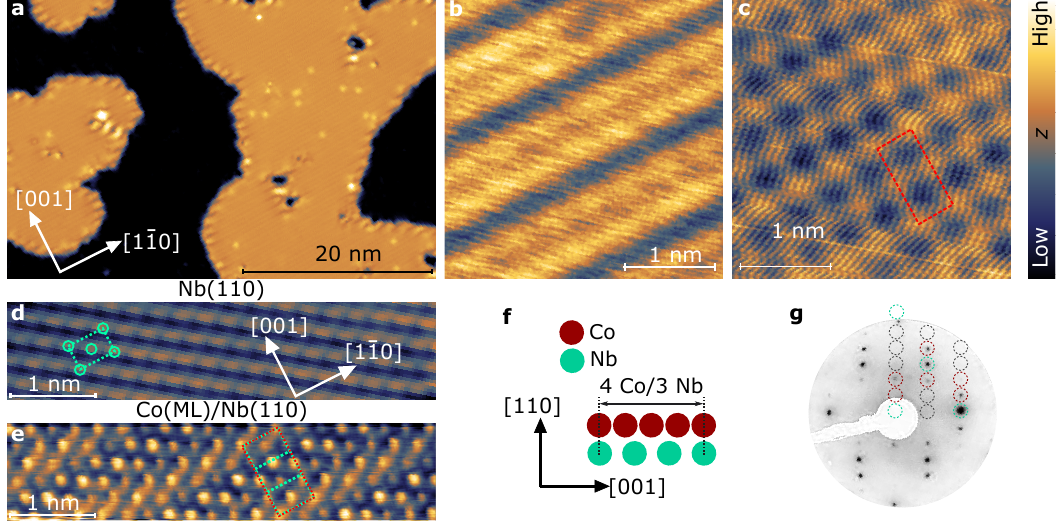}
\caption{Structural characterization of Co monolayer islands on the Nb(110) surface. \textbf{a} Constant-current STM image of a typical sample, where about $50\%$ of a Co monolayer is deposited on a clean and unreconstructed Nb(110) surface. A stripe-like pattern along the $[1\overline{1}0]$ crystallographic direction indicates the presence of a uniaxial reconstruction in the Co islands. \textbf{b} A zoom-in on the Co island showing the stripe-like reconstruction. \textbf{c} A different appearance of the structural reconstruction in the Co monolayer obtained for different tunneling parameters, where the red dotted rectangle indicates the structural unit cell. \textbf{d} Constant-current STM image showing the atomic corrugation of the Nb(110) surface, with the structural unit cell indicated by the cyan dotted rectangle. \textbf{e} Constant-current STM image showing the atomic corrugation of the Co/Nb(110) surface, with its structural unit cell, indicated by the red dotted rectangle, 
compared with the unit cell of the Nb(110) surface underneath, indicated by the cyan dotted rectangle. \textbf{f} Side-view schematic of the $(3\times1)$ reconstruction, with 4 Co atoms sitting on top of 3 Nb atoms along the [001] crystallographic direction. \textbf{g} LEED pattern acquired from the Co/Nb(110) sample with an electron beam energy of 70~eV, showing the primary spots of the Nb(110) surface (highlighted with cyan dotted circles) and the additional spots (highlighted with red dotted circles) due to the $(3\times1)$ reconstruction along the [001] crystallographic direction of the Co monolayer. The remaining gray dotted circles indicate other expected LEED spots due to the $(3\times1)$ reconstruction which were not clearly visible in the LEED pattern. Scanning parameters: \textbf{a} \textit{I}=0.5~nA, \textit{V}=50~mV; \textbf{b} \textit{I}=5~nA, \textit{V}=20~mV; \textbf{c} \textit{I}=20~nA, \textit{V}=20~mV; \textbf{d} \textit{I}=5~nA, \textit{V}=2~mV; \textbf{e} \textit{I}=20~nA, \textit{V}=2~mV. \textit{T}=4.2~K.}\label{fig1}
\end{figure}

The preparation of an atomically clean (110) surface of a Nb single crystal and subsequent deposition of about half a monolayer (ML) of Co (see Methods for more details) results in the formation of Co islands like the ones shown in Fig.~\ref{fig1}a.
In the acquired constant-current STM images, the Co islands appear with a stripe-like pattern along the $[1\overline{1}0]$ crystallographic direction of the Nb(110) surface, indicating the presence of a reconstruction in the Co layer on top of the Nb surface.
The magnified images in Fig.~\ref{fig1}b-c of the reconstructed Co obtained at different bias voltages show the changing appearance of the reconstruction, and the uncovered details allow us to identify the structural unit cell (red dotted rectangle).

In order to compare the unit cell of the reconstructed Co with the one of the Nb(110) surface, STM images with atomic resolution are acquired.
Figure~\ref{fig1}d and~\ref{fig1}e show the atomic corrugation at the bare Nb(110) surface and at the Co surface, respectively.
Figure~\ref{fig1}d reveals the bcc(110) structure of the Nb(110) surface, where the structural unit cell is indicated by the cyan dotted rectangle.
In Fig.~\ref{fig1}e the details of the atomic configuration in the Co are directly visible, clearly showing a compression of the Co layer along the [001] crystallographic direction.
The same unit cell (red dotted rectangle) identified in Fig.~\ref{fig1}c is also observed in the atomically resolved image in Fig.~\ref{fig1}e, where a stripe-like modulation superimposed on the atomic corrugation is clearly visible.
Based on these STM images, we are able to conclude that a $(3\times1)$ reconstruction is present in the Co, with 4 Co atoms sitting on top of 3 Nb atoms along the [001] crystallographic direction of the Nb(110) surface underneath.
This is schematically illustrated in Fig.~\ref{fig1}f.
Furthermore, a low-energy electron diffraction (LEED) pattern acquired from the Co/Nb(110) sample (see Fig.~\ref{fig1}g) confirms the presence of a bcc(110) structure with additional peaks consistent with a $(3\times1)$ reconstruction like the one observed in the STM images.

Based on the experimental observations of the structural configuration of the Co islands on the Nb(110) surface, first-principles density functional theory (DFT) calculations were performed using the Vienna Ab initio Simulation Package (VASP)~\cite{Kresse1996,Kresse1996a,Hafner2008} (see Appendix's for more details) in order to determine the structural ground state, and in particular the adsorption sites of the Co atoms on the Nb(110) surface.
A $(3\times1)$ reconstruction with 8 Co atoms for 6 Nb atoms in the structural unit cell was considered based on the experimental data.
The relaxed geometry is reported in Fig.~\ref{fig2}a, while the atomic positions are reported in Appendix~\ref{sec:reconstruction} and in Fig.~\ref{figS3}. 
The reconstructed Co unit cell is highlighted by a red dashed rectangle.
The inset shows a simulated constant-current STM image of the reconstructed Co monolayer on the Nb(110) surface based on the DFT calculations using the BSKAN code~\cite{HoferPSS2003,MandiPRB2015}. The calculated STM image closely resembles the constant-current STM image in Fig.~\ref{fig1}c, demonstrating the agreement between simulations and experiments.
Overall, the Co monolayer nearly assumes the triangular atomic structure of a single layer of bulk hcp Co, with only slight rearrangements due to the Nb(110) surface below.
The reconstructed geometry preserves the twofold rotational symmetry of the Nb(110) surface, as visible from the measured and the simulated STM images in Figs.~\ref{fig1}b-e and \ref{fig2}a.

We use the calculated atomic structure to determine atomistic spin-model parameters via the relativistic torque method~\cite{Udvardi2003} as implemented within the screened Korringa--Kohn--Rostoker code~\cite{Szunyogh1995}.
The magnetic ground state is determined from spin-dynamics simulations based on this spin model.
The result is the prediction of an in-plane ferromagnetic ground state for the Co monolayer, with a magnetic easy axis along the $[001]$ crystallographic direction of the Nb(110) surface, preferred by about 0.013~meV/atom with respect to the $[1\overline{1}0]$ and by 0.304~meV/atom with respect to the out-of-plane $[110]$ direction, as illustrated in Fig.~\ref{fig2}b, where the white arrows indicate the atomic magnetic moments in the Co monolayer unit cell.

\begin{figure}[ht!]
\centering
\includegraphics[width=0.5\linewidth]{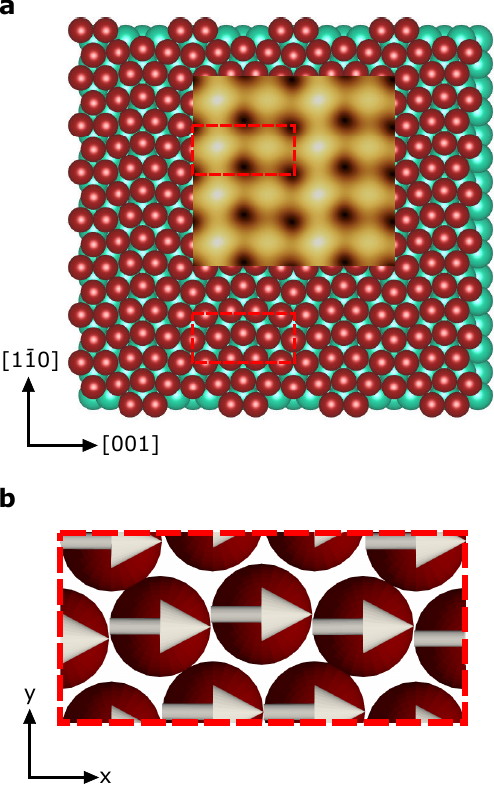}
\caption{Structural and magnetic model from first-principles calculations. \textbf{a} Relaxed atomic structure for the Co monolayer on the Nb(110) surface. The structural unit cell is indicated by the red dotted rectangle, which is in agreement with the atomic configuration observed experimentally; see Fig.~\ref{fig1}c. The inset shows a simulated constant-current STM image of the calculated Co/Nb(110) atomic arrangement, which reproduces closely the experimentally acquired image shown in Fig.~\ref{fig1}e. \textbf{b} In-plane ferromagnetic ground state of the Co monolayer obtained from first-principles-parametrized spin-model simulations, with the magnetic easy axis \textit{x} and perpendicular axis \textit{y} along the $[001]$ and $[1\overline{1}0]$ crystallographic directions, respectively.}\label{fig2}
\end{figure}

\subsection{Modeling}\label{modeling}
\begin{figure}[ht!]
\centering
    \includegraphics[width=\textwidth]{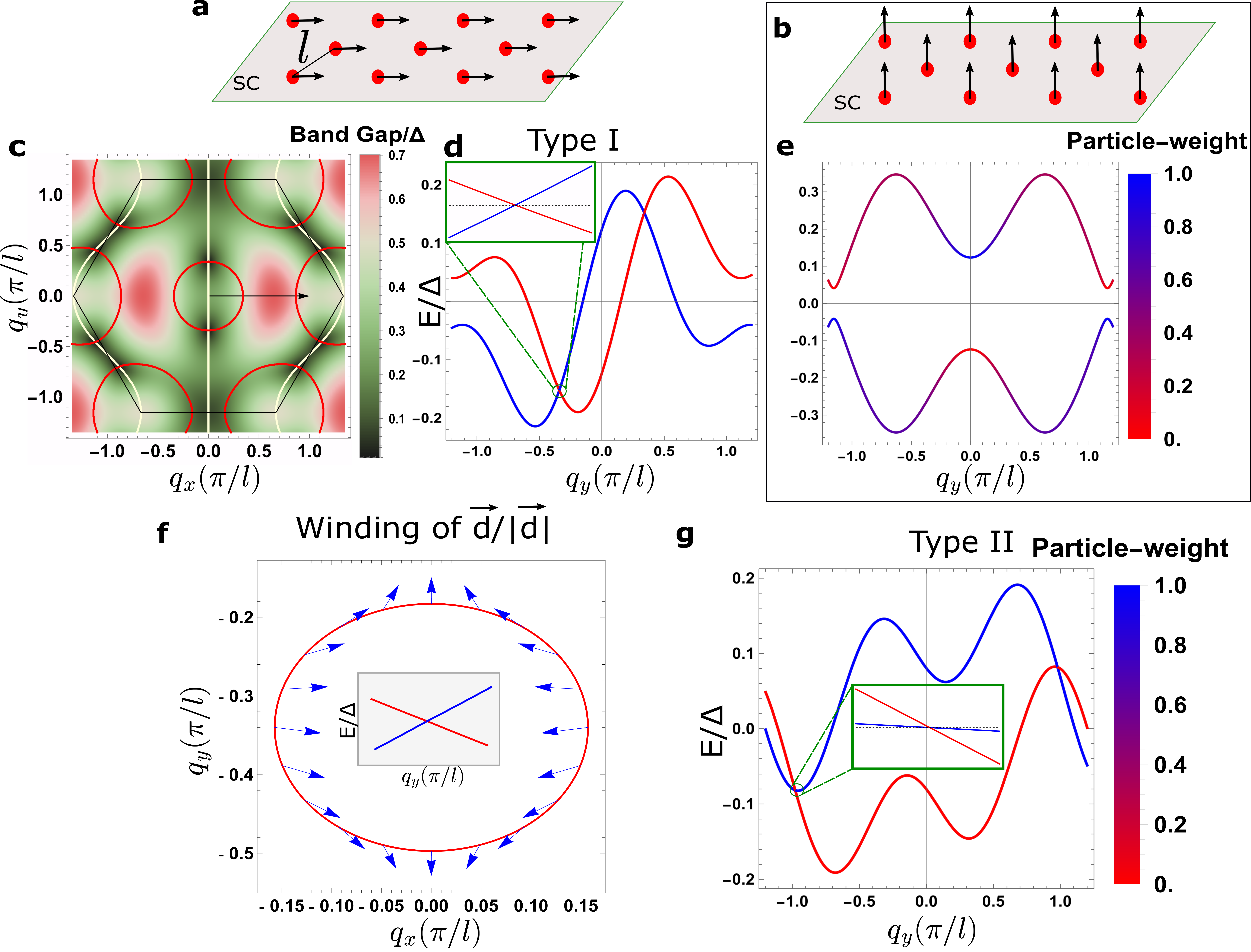}
\caption{Formation of type I and II Weyl cones with non-vanishing winding number in the low-energy theory.  \textbf{a} Schematics for in-plane magnetization. \textbf{b} Schematics for out-of-plane magnetization. \textbf{c} Energy difference between upper and lower band $\delta E=E_{+}-E_{-}$, plotted in the hexagonal BZ with the magnetization direction (black arrow) along the $x$ direction ($[001]$). The intersection of the nodal lines of the $d_y$ (light yellow) and $d_z$ (red) components of the effective Hamiltonian in Eq.~(\ref{eq:GiannisEffHam}) coincide with the vanishing of the band gap (dark green on the colormap). \textbf{d} Energy bands plotted in a linecut along the $q_y$ direction, which shows the appearance of a pair of tilted type I Weyl cones (gray inset) for an in-plane magnetization, $\zeta=\pi/2$. \textbf{e} Same as in $\textbf{d}$ but for the out-of-plane case, $\zeta=0$. \textbf{f} Winding of $\vec{d}/|\vec{d}|$ around the Weyl point shown in \textbf{d}, indicating the topological charge. \textbf{g} Appearance of type II Weyl nodes for larger spin-orbit coupling than in \textbf{d}.
In all plots, we use a next-nearest-neighbor model, and the parameters are fixed at $m=10\hbar^2 l^{-2}/\Delta$, $\mu=10\Delta$, $a=0.5l\cdot\Delta $, $J=4\upsilon_\mathrm{F}l^{-2}/\left(k_\mathrm{F+}+k_\mathrm{F-}\right)$ (to ensure that $\epsilon_{\pm}\rightarrow0$), except for panel \textbf{g} where $a=7 l\cdot\Delta$ is used.}
\label{fig5}
\end{figure}

Motivated by the experimental results shown in Fig.~\ref{fig1} and the findings of the first-principles calculations presented in Fig.~\ref{fig2}, we develop the single-particle low-energy theory of a triangular array of magnetic adatoms with in-plane magnetic moments on top of a SC substrate (see Fig.~\ref{fig5} for details).
To this end, we approximate the Nb(110) surface by a continuum two-dimensional bulk superconductor and write the microscopic Hamiltonian
\begin{equation}\label{eq:GiannisMicroHam}
\begin{split}
    &\hat{H}=\hat{H}_{\rm SC}+\hat{H}_{\rm Ex}\ \textrm{where}, \\
    &\hat{H}_{\rm SC}= \dfrac{1}{2}\sum_{\vec{k}} \Psi^{\dagger}_{\vec{k}}\left(\tau_z \left( \xi_{\textbf{k}}+a\left( k_y \sigma_x-k_x\sigma_y\right)\right)+\Delta \tau_x\right)\Psi_{\vec{k}},\\
     &\hat{H}_{\rm Ex}= -\dfrac{J}{2} \sum_{\vec{k},\vec{k}',j} e^{i (\vec{k}'-\vec{k})\vec{r}_j}\cdot \Psi^{\dagger}_{\vec{k}'} \left(\Vec{S}_j \cdot \Vec{\sigma}\right) \Psi_{\vec{k}},  
\end{split}
\end{equation}
are the Hamiltonians describing the superconducting substrate ($H_{\rm SC}$) with conventional Rashba spin-orbit coupling and pairing of strengths $a$ and $\Delta$, respectively, and the exchange interaction of the magnetic adatoms with the substrate electrons ($H_{\rm Ex}$) of strength $J$, both defined in the Nambu basis $\Psi_{\vec{k}}=\begin{pmatrix}\psi_{\vec{k}\uparrow},\psi_{\vec{k}\downarrow},\psi^{\dagger}_{-\vec{k}\downarrow},-\psi^{\dagger}_{-\vec{k}\uparrow}\end{pmatrix}^T$.
Here, the Pauli matrices $\tau$ and $\sigma$ act on particle-hole and spin space, respectively, and the operators $\psi_{\vec{k}s}$ annihilate electrons with wave vector $\vec{k}$ and spin $s$. 
Also, we consider a quadratic dispersion in the superconductor $\xi_{\vec{k}}=\hbar^2\vec{k}^2/2m-\mu$, where $\hbar$ is the reduced Planck constant, $m$ is the electron mass, $\mu$ is the chemical potential, and classical uniform magnetic moments $\Vec{S}_j=S \begin{pmatrix}  \cos(\theta)\sin(\zeta),\sin(\theta)\sin(\zeta),\cos(\zeta) \end{pmatrix}$ at positions $\vec{r}_j$, where $\theta$ and $\zeta$ are the azimuthal and polar angles of the magnetization vector, respectively.
Consistent with the experimental geometry of the Co monolayer and first-principles calculations, see Fig.~\ref{fig1} and Fig.~\ref{fig2}, we consider the magnetic adatom position vectors $\vec{r}_j$ to cover a triangular lattice.
In the case where the indirect couplings between magnetic adatoms in different sites are negligible, the term $\hat{H}_{\rm Ex}$ in Eq.~(\ref{eq:GiannisMicroHam}) induces pairs of bare YSR states with single-particle energies $\epsilon_{\pm}=\pm\Delta\left( 1-(\pi \nu JS )^2\right)/\left(1+(\pi \nu JS )^2\right)$, where $\nu$ is the normal-state density of states at the Fermi level.
This Hamiltonian reflects the twofold rotational symmetry of the experimental system around the $z$ axis, which together with an in-plane magnetization results in an effective time-reversal symmetry for the system. 

Following a similar derivation as in Refs.~\cite{Li2016,PhysRevB.88.155420,PhysRevB.91.064505}, we project the full Hamiltonian in Eq.~(\ref{eq:GiannisMicroHam}) to the bare YSR state wave functions in the limit $\epsilon_{\pm}\rightarrow0$, and obtain an effective Hamiltonian by considering the leading-order terms in $E/\Delta$ and a dilute lattice of magnetic adatoms $k_{\textrm{F}} l\gg1$, where $k_{\textrm{F}}$ is the Fermi wave vector of the bulk and $l$ the lattice spacing of the 2D adatoms lattice (see Appendix~\ref{sec:Low-EnergyTheory} for details).
In the basis of the bare YSR states at each adatom site, we assume periodic boundary conditions and write the two-by-two matrix block of the effective Hamiltonian in Fourier space as

\begin{equation}\label{eq:GiannisEffHam}
\begin{split}
    \mathcal{H}(\Vec{q})=&d_0(\Vec{q})I+ \Vec{d}(\Vec{q})\cdot \Vec{\sigma},
   \\ d_0(\Vec{q})=&\dfrac{iJ}{2} \sin(\zeta) \sum_{\Vec{r} \neq 0} e^{-i\Vec{q} \Vec{r}}s_0(r) \sin(\phi(\Vec{r})-\theta), \\ 
    d_x(\Vec{q})=&\dfrac{-iJ}{2} \cos(\zeta) \sum_{\Vec{r} \neq 0} e^{-i\Vec{q} \Vec{r}}s_y(r) \sin(\phi(\Vec{r})-\theta),\\
    d_y(\Vec{q})=&\dfrac{iJ}{2} \sum_{\Vec{r} \neq 0} e^{-i\Vec{q} \Vec{r}}s_y(r) \cos(\phi(\Vec{r})-\theta), \\  d_z(\Vec{q})=&-\dfrac{J}{2} \sum_{\Vec{r} \neq 0} e^{-i\Vec{q} \Vec{r}}s_z(r),
\end{split}
\end{equation}
where $I$ is the unit matrix, the radial parts $s_i(r)$ are functions that depend on the microscopic characteristics of the bulk superconductor and the approximations used, see 
Appendix~\ref{sec:Low-EnergyTheory}, and the sums run over all positions of the lattice $\vec{r}$ with in-plane angle $\phi(\vec{r})$.
The spectrum of $\mathcal{H}(\Vec{q})$ is 
\begin{equation} \label{eq:GiannisSpectrum}
E_{\pm}(\Vec{q})=d_0(\Vec{q})\pm\sqrt{  d^2_x(\Vec{q})+d^2_y(\Vec{q})+ d^2_z(\Vec{q})}.
\end{equation}
For a generic magnetization direction with a finite out-of-plane magnetization component and non-vanishing spin-orbit coupling, the spectrum is gapped.
Based on the structural symmetries of the experimental setup and the results of the first-principles calculations in Sec.~\ref{struANDmag}, we focus on the case of an in-plane magnetic structure $\zeta=\pi/2$ and $\theta=0$ (see  Fig.~\ref{fig5}a).
In this case, the model preserves the effective time-reversal symmetry of the original system represented as $\mathcal{T}_{\textrm{eff}}=\sigma_{z}\mathcal{K}$, where $\mathcal{K}$ denotes complex conjugation.
Note that $\mathcal{T}_{\textrm{eff}}$ does not change the wave vector $\Vec{q}$, since it represents a combination of physical time reversal and twofold rotation both reversing the in-plane wave vectors.
Together with the particle-hole constraint $\mathcal{C}=\sigma_{x}\mathcal{K}$, which transforms $\Vec{q}$ into $-\Vec{q}$, this requires the coefficient $d_{x}(\Vec{q})$ to vanish.
Here, the spectrum in Eq.~(\ref{eq:GiannisSpectrum}) reduces to $E_{\pm}(\Vec{q})=d_0(\Vec{q})\pm\sqrt{ d^2_y(\Vec{q})+ d^2_z(\Vec{q})}$, and the spectral gap vanishes when $E_{+}(\Vec{q}_0)=E_{-}(\Vec{q}_0)$ for the momenta $\Vec{q}_0$ where tilted Weyl cones form, see Fig.~\ref{fig5}d.
Therefore, locating the Weyl cones is equivalent to finding the points where the nodal lines $d_y(\Vec{q})=0$ and $d_z(\Vec{q})=0$ intersect, see Fig.~\ref{fig5}c.
While the nodal lines of $d_y(\Vec{q})$ (light yellow lines) explicitly depend on the in-plane magnetization angle $\theta$, the nodal lines of $d_z(\Vec{q})$ (red lines) remain invariant. 

In Figs.~\ref{fig5}c,d,f and g, we fix the in-plane magnetization orientation to the direction determined from the first-principles calculations.
We note the appearance of multiple pairs of Weyl cones in the Brillouin zone, which is linked to the long-range bulk-mediated couplings between YSR states.
We confirm the topological origin of the Weyl nodes by calculating their topological charges, related to the winding of the unit vector $\vec{d}/|\vec{d}|$ in the neighborhood of the band crossings \cite{sato2017TopologicalSuperconductorsReview}, see Fig.~\ref{fig5}f.
The Weyl nodes appear in pairs with opposite topological charge at $\Vec{q}$ and $-\Vec{q}$ wave vectors in the Brillouin zone.
The tilting is caused by the $d_0(\Vec{q})$ term in the Hamiltonian, which switches sign upon reversing the wave vector $\Vec{q}$.
Depending on the amount of tilting, the Weyl nodes can be classified either as type I or type II. 
For type I nodes, one of the cones starting from the nodal point contains only states which have a higher energy than the state at the nodal point, while the other cone only contains states with a lower energy, see the inset of Fig.~\ref{fig5}d.
For type II nodes, the Weyl cones intersect the energy level of the node due to the high amount of tilting, see the inset of Fig.~\ref{fig5}g.
Whether type I or type II nodes are present depends on the system parameters, see Figs.~\ref{fig5}d and g.
For a more detailed discussion on this classification, see Appendix \ref{sec:Low-EnergyTheory}.
The $d_0(\Vec{q})$ and $d_y(\Vec{q})$ terms are only finite in the presence of spin-orbit coupling, which hence is required for the formation of Weyl nodes out of the nodal lines of $d_z(\Vec{q})=0$, and the tilting of these nodes.
In contrast to the in-plane case, an out-of-plane component in the magnetization in Fig.~\ref{fig5}b leads to a non-vanishing $d_x$ component in Eq.~(\ref{eq:GiannisSpectrum}), which ultimately gaps the Weyl cones, see Fig.~\ref{fig5}e.
In the case when the magnetization is completely out-of-plane (see Fig.~\ref{fig5}b), the bulk spectrum acquires a mini-gap since the tilting component $d_0(\Vec{q})$ vanishes.

\subsection{Spectroscopic characterization}\label{spectroscopy}
\begin{figure}[ht!]
\centering
\includegraphics[width=1\linewidth]{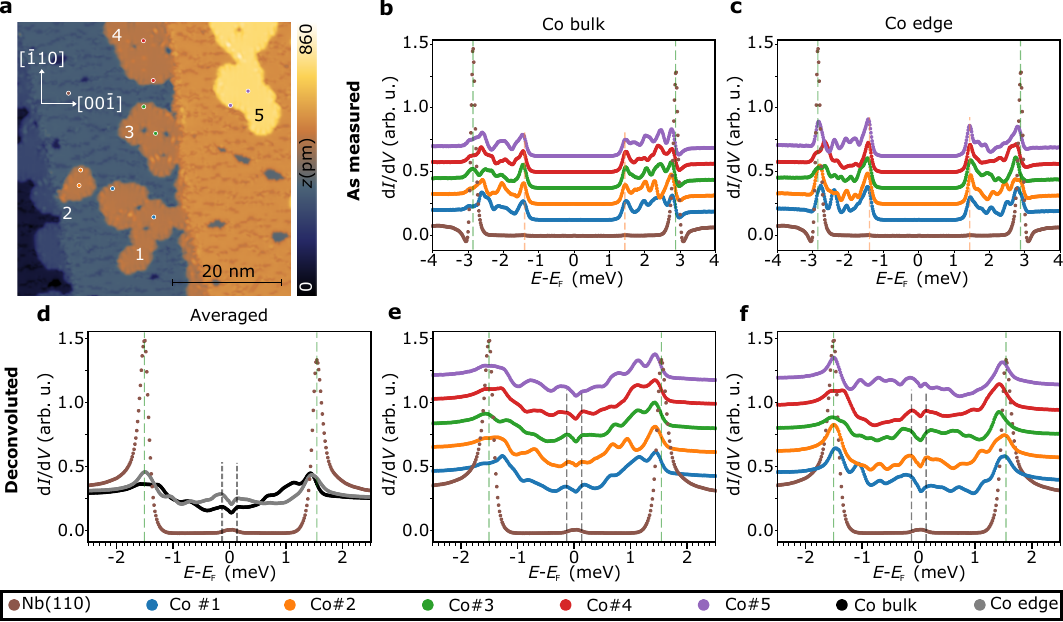}
\caption{Scanning tunneling spectroscopy at Co monolayer islands on the Nb(110) surface. \textbf{a} Constant-current STM image of the Co islands labeled $1$ to $5$ on the Nb(110) surface (scanning parameters: \textit{I}=200~pA, \textit{V}=10~mV).
The dark patches visible on the Nb(110) terraces are oxygen contamination aggregates remaining after the cleaning process.  \textbf{b}-\textbf{c} As-measured d\textit{I}/d\textit{V} spectra on Co islands acquired in the middle of the islands \textbf{b} and on the edge of the islands \textbf{c}. Spectra are color-coded to specific islands whose positions are shown in panel \textbf{a}. In all plots, the brown graphs mark the substrate spectrum for reference. The value of $\Delta\textsubscript{tip}$ is marked by an orange dashed line, while the green dashed line shows the value of $\Delta\textsubscript{tip}+\Delta\textsubscript{Nb(110)}.$ \textbf{d}-\textbf{f} LDOS obtained via deconvolution of d\textit{I}/d\textit{V} shown in panels \textbf{b} and \textbf{c}. \textbf{d} Data averaged over all islands for the middle of the islands (in black) and the edges (in gray). In panels \textbf{e} and \textbf{f} spectra for the middle and the edge of each island are shown, respectively. In panels b,c, e, and f the curves are offset by 0.15 arb. u. for clarity. Stabilization parameters: \textit{I}$_{\textrm{stab}}$=1~nA; \textit{V}$_{\textrm{stab}}$=4~mV; $\Delta{V}$=16~$\mu$V. d\textit{I}/d\textit{V} spectra acquired with a superconducting tip with $\Delta\textsubscript{tip}$=1.43~meV at a measurement temperature T=300~mK.}\label{fig3}
\end{figure}

The presence of an in-plane ferromagnetic ground state in the Co monolayer on Nb(110) provides an opportunity to experimentally test the above model by examining the in-gap local density of states (LDOS) of the sample in its superconducting state.
A new sample was prepared and cooled down to $T=300$ mK, to establish a well-developed superconducting phase in the Nb substrate ($T_{\textrm{c}}=9.2$ K) and enable scanning tunneling spectroscopy measurements with high energy resolution.
In addition, a superconducting Nb tip is employed in order to characterize the LDOS of the prepared samples with enhanced energy resolution (smaller than 100 $\mu$eV~\cite{WiebeRSI2004}, see Methods section).
Figure~\ref{fig3}a shows a constant-current STM image of the prepared sample, where five Co islands indicated by the numbers 1 to 5 are visible.
Figures~\ref{fig3}b and c show the as-measured spectra at specific locations of the five Co islands presented in Fig.~\ref{fig3}a. Figures~\ref{fig3}d-f report the LDOS as obtained from the deconvolution of the measured d\textit{I}/d\textit{V} presented in panels b and c (see Methods section for more details).
The brown curve is the LDOS at the bare Nb(110) surface, showing the expected hard superconducting gap with coherence peaks at $\pm\Delta_{\textrm{Nb}}=\pm1.5$ meV.
The curves are color-coded to specific locations marked in Fig.~\ref{fig3}a, while the black and gray curves in Fig.~\ref{fig3}d show the averaged data from all five islands for the centre and the edge, respectively.
LDOS curves both at the center and at the edge of the Co islands show the presence of in-gap states at all energies between -1.5 meV and +1.5 meV.
This observation experimentally confirms that the Co atoms have magnetic moments which couple, via exchange interaction, to the superconducting Nb substrate.
While the details of the different spectra differ, we consistently see a double-peak feature around the Fermi level, $E=E_{F}$, with peak positions at about $E=\pm100$ $\mu$eV (see gray dashed lines in Fig.~\ref{fig3}d-f).
On all islands, we observe the highest amplitude of this double-peak feature at the edges, although it remains visible at all positions.
Finally, the experimentally determined LDOS in Figs.~\ref{fig3}d-f is enhanced around $E=E_{F}$ at the edges of the Co islands, compared to the islands' centre.

\begin{figure}[ht!]
\centering
\includegraphics[width=1\linewidth]{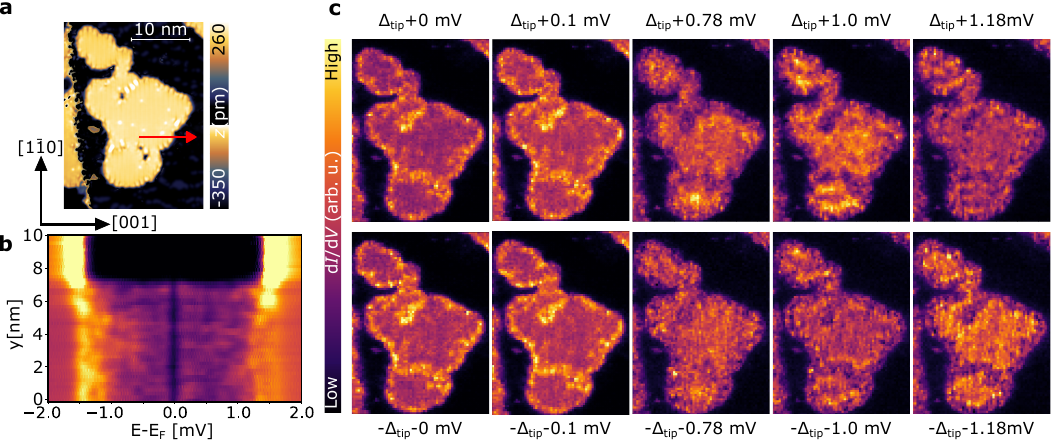}
\caption{Spatially resolved local density of states of a Co monolayer island on the Nb(110) surface. \textbf{a} Constant-current STM image of a Co island on the Nb(110) surface. \textbf{b} Waterfall plot of the LDOS obtained by deconvolution of differential tunneling conductance spectra acquired over the edge of a Co island marked by a red arrow in panel \textbf{a}.
Stabilization parameters: \textit{I}= 1~nA; \textit{V}= 4~mV; $\Delta{V}$=16~$\mu$V. \textbf{c} Differential tunneling conductance maps of the same Co island in panel \textbf{a} acquired at different bias voltages \textit{V}. Stabilization parameters: \textit{I}$_{\textrm{stab}}$=1~nA; \textit{V}$_{\textrm{stab}}$=-4~mV; $\Delta{V}$=40~$\mu$V. d\textit{I}/d\textit{V} maps and spectra are acquired with a superconducting tip with $\Delta\textsubscript{tip}$=1.43 meV. Scanning parameters: \textit{I}= 200~pA, \textit{V}= -10~mV.}\label{fig4}
\end{figure}

To access the details of the spatial dependence of the LDOS in the Co islands, d\textit{I}/d\textit{V} spectroscopy maps were acquired at several energies inside the superconducting gap of the Nb substrate for a specific magnetic island, as collected in Fig.~\ref{fig4}.
Figure~\ref{fig4}a shows the topography of the Co island used for the LDOS($E$) mapping.
The red arrow indicates the line along which the d\textit{I}/d\textit{V} spectra shown in Fig.~\ref{fig4}b as waterfall plots were acquired, which reveal the spatial evolution of the LDOS along a line starting at the Co island, crossing its edge and ending onto the Nb(110) surface.
The obtained LDOS spatial evolution is consistent with the experimental observations for other Co islands reported in Fig.~\ref{fig3}, showing a double-peak feature around $E_{\textrm{F}}$ that is enhanced near the edge of the Co island. 
Figure~\ref{fig4}c shows d\textit{I}/d\textit{V} maps acquired at several biases in the intervals $-|(\Delta\textsubscript{sample}+\Delta\textsubscript{tip})/e|<V<-|\Delta\textsubscript{tip}/e|$ and $+|\Delta\textsubscript{tip}/e|<V<+|(\Delta\textsubscript{sample}+\Delta\textsubscript{tip})/e|$, in order to characterize the low-energy LDOS of the Co island.
First, for bias values close to the Fermi level, $V\textsubscript{F}=\pm|\Delta\textsubscript{tip}/e|$, the d\textit{I}/d\textit{V} maps in Fig.~\ref{fig4}c show an enhancement of the LDOS at all edges of the Co island. This LDOS enhancement at the Fermi level is not only observed at the edge, but also at topographic imperfections in the interior of the Co island, such as holes and dislocation lines.
Second, for biases $V-|\Delta\textsubscript{tip}/e|\approx\pm100~\mu{V}$, the d\textit{I}/d\textit{V} maps show a maximal LDOS at the edges, confirming the enhancement of the double-peak feature around the island's edges, as already observed in the waterfall plot in Fig~\ref{fig4}b and in the point spectra reported in Fig.~\ref{fig3}.
Third, at larger biases, $\Delta\textsubscript{tip}/e+0.78$ mV $<|V|<$ $\Delta\textsubscript{tip}/e+1.18$ mV), the LDOS in the interior of the Co island is larger than at the edge, indicating that in this energy range, the LDOS is mostly originating from \textit{bulk} states located inside the Co island.
Interestingly, in this energy range, we can also observe a quasiparticle interference pattern which is the most pronounced in the lower part of the magnetic island, where the emergent electronic states in the bulk of the two-dimensional Co layer are confined by the presence of the island's edge at the bottom and a line of topographic imperfections 5-6 nm away.
In this specific area, single, double, and triple maxima along the $[1\overline{1}0]$ direction are observed at $\Delta\textsubscript{tip}/e+0.78$ mV, $\Delta\textsubscript{tip}/e+1.00$ mV, and $\Delta\textsubscript{tip}/e+1.18$ mV, respectively.

\subsection{Real-space simulations}

\begin{figure}[ht!]
\centering
\includegraphics[width=1\linewidth]{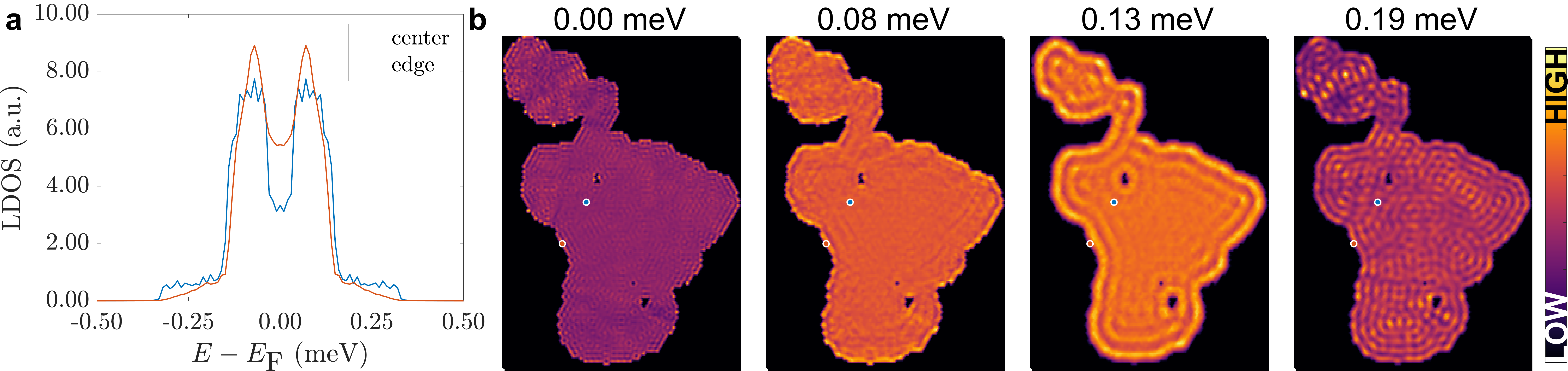}
\caption{
    Calculated spectrum of a Co monolayer island.
    \textbf{a} Calculated LDOS inside (blue) and at the edge of (orange) the island. The positions of the atoms are shown as correspondingly colored dots in panel \textbf{b}.
    \textbf{b} Two-dimensional LDOS map at the energies $E-E_{\textrm{F}}$ indicated above the panels.
    Parameters used for the simulations in Eq.~\eqref{eq:TBHam} are $t_{1}=41.7~\mu$eV, $t_{2}=16.7~\mu$eV, $\mu+J=30.0~\mu$eV, $\alpha_{1}=-3.33~\mu$eV, $\alpha_{2}=8.33~\mu$eV, $\Delta=100~\mu$eV, and $J=1000~\mu$eV. 
}\label{fig6}
\end{figure}

To elucidate the difference in the experimental spectra obtained at the centers and at the edges of the islands theoretically, we consider a lattice version of the continuum theory in Eq.~\eqref{eq:GiannisMicroHam} including nearest-neighbor and next-nearest-neighbor hopping terms, see Methods.
It is worth noting that the parameters in the model are chosen to reproduce the observed spectral features very close to the Fermi level $\left|E-E_{\textrm{F}}\right|<0.1$~meV, not the whole range of in-gap states or the full band structure of bulk Nb. The simulated spectral properties are shown in Fig.~\ref{fig6}.

As it can be seen in Fig.~\ref{fig6}a, the pair of peaks at around $E-E_{\textrm{F}}=\pm100~\mu$eV is reproduced by the real-space model, together with the decrease in the LDOS at the Fermi level.
Moreover, no induced gap is observed, in agreement with the low-energy model for MSHs with in-plane magnetization discussed above.
The LDOS at the edge of the island is enhanced for $\left|E-E_{\textrm{F}}\right|<0.1$~meV, as visible in the first two panels of Fig.~\ref{fig6}b.
With the chosen set of parameters, we only observed this enhancement at the edges in the presence of spin-orbit coupling.
The spin-orbit coupling is responsible for the formation of Weyl nodes and their tilting as well.
However, edge states connecting the Weyl nodes are not visible in the projection of the Brillouin zone on the edge direction; see Appendix~\ref{sec:simulations} and Fig.~\ref{figS1} for details.
At higher energies, the LDOS at the edge becomes lower than in the interior of the island, similarly to the experimental trend, although this is observed in the experiments further away from the Fermi level than in the model. 
The real-space simulations also reveal the formation of modulations of the LDOS close to the edges at $E-E_{\textrm{F}}=0.13$~meV and at $E-E_{\textrm{F}}=0.19$~meV, which are qualitatively similar to the confined oscillations between $E-E_{\textrm{F}}=0.78$~meV and $E-E_{\textrm{F}}=1.18$~meV in Fig.~\ref{fig4}c.

\section{Discussion}\label{discussion}
We have investigated the physics of in-plane ferromagnetic 2D MSH systems, experimentally realized in the model system consisting in a magnetic Co monolayer on a superconducting Nb(110) single crystal.
In contrast to the gapped superconducting phase found for out-of-plane ferromagnetic systems, the effective time-reversal symmetry of in-plane systems leads to the formation of protected nodal points, as derived by our analytical low-energy model and tight-binding calculations.
The nodal points are shifted away from the Fermi level in opposite directions at opposite wave vectors, resulting in the tilting of the Weyl cones of the emergent electronic band structure.
The predicted gapless nodal band structure is consistent with the fact that the experimental data offer no indication of a gapped superconducting phase and clearly show a double-peak feature in the LDOS around the Fermi level.
This double-peak spectroscopic feature is strongest at the edges of the investigated Co monolayer islands, an observation which can be reproduced by our real-space model as well. 
It is worth mentioning that the dislocation lines and other imperfections inside the Co island show similar spectroscopic signatures as the edges, suggesting the possibility that our observations could potentially be explained as non-topological in nature. 
Yet, with a nodal character of the in-gap band structure, such universal spectroscopic signatures are expected. The imperfections reshape the topologically non-trivial region, which results in additional topologically non-trivial edges, generating the edge intensity observed experimentally. 

Furthermore, it is worth noting that the spectroscopic signatures of the topological superconducting phase may also differ from system to system.
In certain parameter regimes, our models predict a large asymmetry in the spectrum along different crystallographic edges, depending on the orientation of the edge with respect to the magnetization direction.
In order to investigate the subtle spectroscopic difference of magnetic edges along different crystallographic directions, similarly to what was done previously for the antiferromagnetic Mn monolayer on Nb(110)~\cite{Bazarnik2023-natcomm}, one would need to prepare an in-plane ferromagnet/superconductor system which would show a pseudomorphic growth with straight edges, or to craft a 2D spin lattice via STM-based atom manipulation, similarly to what was done for 2D Cr lattices on Nb(110)~\cite{SoldiniNAT-PHYS2023}.
A third option would be to employ van der Waals magnet-superconductor heterostructures, where the growth of the magnetic layer is less bound to follow the atomic structure of the surface of the substrate underneath.
This was recently attempted by Cuperus et al.~\cite{Cuperus2025}, even though no evidence of topological superconductivity was observed due to a weak exchange coupling between the magnetic and the superconducting van der Waals layers.
Our experimental investigation shows a very strong exchange coupling between the Co in-plane magnetic layer and the Nb(110) superconducting substrate.
Those observations propose the combination of van der Waals and non-van der Waals materials as a potentially promising approach for the study of tilted Weyl cones in in-plane magnetized systems and the influence of the magnetization direction on the spectral features at different edges.

Finally, the developed low-energy analytical model demonstrates that the Weyl cones can be either of type I or type II depending on the details of the chosen parameter values.
Accordingly, the present work proposes in-plane ferromagnet-superconductor hybrids as a promising materials platform for the exploration of the unique transport properties of superconducting Weyl systems.

\section{Methods}\label{Methods}

\subsection{Sample preparation}
A clean Nb(110) surface was obtained by repeatedly flashing the single crystal to \textit{T}\textgreater2,700~K. 
Monolayer thick Co islands were grown on top of an unreconstructed Nb(110) single crystal surface via molecular beam epitaxy by e-beam evaporation from a Co rod under ultra-high vacuum conditions with a base pressure of about $1.0\times10^{-10}$ mbar. A bulk Cr tip is used for STM measurements at 4.2 K, whereas a bulk Nb tip is used for the STS experiments at 300 mK.

\subsection{Scanning tunneling microscopy and spectroscopy experiments}
All experiments were performed in two home-built low-temperature scanning tunneling microscopes operating at a base temperature of \textit{T}~=~4.2~K and \textit{T}~=~300~mK~\cite{WiebeRSI2004}, respectively.
STM images were obtained in constant-current mode. For the differential conductance (d$I$/d$V$($V$)) spectra, the tip was stabilized at bias voltage \textit{V}$_{\textrm{stab}}$ and current \textit{I}$_{\textrm{stab}}$.
Subsequently, the feedback loop was opened and the bias voltage was swept from \textminus4~mV to +4~mV. The differential tunneling conductance d$I$/d$V$ was measured using a standard lock-in technique with a small modulation voltage \textit{V}$_{\textrm{mod}}$ (r.m.s.) of 80 $\mu$V with modulation frequency $f=4.142$~kHz added to the bias voltage.
The d$I$/d$V$ maps were acquired by recording d$I$/d$V$ spectra on a real-space grid in a \textminus3.2~mV to +3.2~mV range.

\subsection{Deconvolution of superconductor-vacuum-superconductor spectra}

Because we used a bulk Nb tip to carry out the STS experiments, the acquired d$I$/d$V$ spectra do not directly correspond to the LDOS of the sample, as for the case of normal metallic tips, but to the convolution of the density of states of the sample and of the superconducting tip. 
Accordingly, to retrieve the actual LDOS of the sample, we perform a well-established numerical deconvolution of the measured d$I$/d$V$ spectra. 
The tunneling conductance can be expressed as

\begin{equation}\label{eq:Deconv1}
\begin{split}
    \frac{\textrm{d}I}{\textrm{d}V}\left(V,T\right)\propto \int^{+\infty }_{-\infty }{{\rho }_{\mathrm{S}}}\left(E\right)\frac{\partial {\rho }_{\mathrm{T}}\left(E-eV\right)}{\partial V}\left[f\left(E-eV,\ T\right)-f\left(E,T\right)\right]\textrm{d}E\\
    +\int^{+\infty }_{-\infty }{{\rho }_{\mathrm{S}}\left(E\right)}{\rho }_{\mathrm{T}}\left(E-eV\right)\frac{\partial f\left(E-eV,T\right)}{\partial V}\textrm{d}E,
    \end{split}
\end{equation}
where \textit{$\rho$}${}_{\mathrm{S}}$(\textit{E}) is the energy-dependent LDOS of the sample, \textit{$\rho$}${}_{\mathrm{T}}$(\textit{E}) is the LDOS at the tip, \textit{f}(\textit{E,T}) is the Fermi-Dirac distribution function, \textit{T} is the experimental temperature, and \textit{V} is the applied bias between tip and sample.

We discretize the integral in Eq.~\eqref{eq:Deconv1} into \textit{N} points in the interval $V\in [-3.2, +3.2]$ meV. Treating the measured d\textit{I}/d\textit{V} spectrum as a column vector [d\textit{I}/d\textit{V}] of length \textit{N}, Eq.~\eqref{eq:Deconv1} can be expressed as a multiplication of an (unknown) sample LDOS vector [\textit{$\rho$}${}_{\mathrm{S}}$] of length \textit{N} with an \textit{N} x \textit{N} matrix {\textit{A}}
\begin{equation}\label{eq:Deconv2}
    \left[\frac{\mathrm{d}I}{\mathrm{d}V}\right]=\boldsymbol{A}\left[{\rho }_S\right].
\end{equation}
Accordingly, the matrix elements of {\textit{A}} are
\begin{equation}\label{eq:Deconv3}
\begin{split}
    A_{ij}=\left(\frac{\partial {\rho }_{\textrm{T}}\left(E_j-eV_i\right)}{\partial V}\left[f\left(E_j-eV_i,\ T\right)-f\left(E_j,T\right)\right]\right.\\
    \left.+{\rho }_{\textrm{T}}\left(E_j-eV_i\right)\frac{\partial f\left(E_j-eV_i,T\right)}{\partial V}\right)\times \delta E,
\end{split}
\end{equation}
where \textit{E}${}_{j}$ are the discrete energy values and \textit{V}${}_{i}$ the bias voltage values of the experimental tunneling spectrum.
Numerically computing a matrix inverse {\textit{A}}${}^{-1}$ 
and multiplying the result with [d\textit{I}/d\textit{V}] yields an approximate sample LDOS
\begin{equation}\label{eq:Deconv4}
    \left[{\rho }_{\mathrm{S}}\right]\sim {\boldsymbol{A}}^{-1}\left[\frac{dI}{dV}\right]
\end{equation}
The DOS of the superconducting Nb tip is assumed to be a BCS DOS with a phenomenological broadening parameter $\Gamma$ following Dynes et al.~\cite{DynesPRL1978}:
\begin{equation}\label{eq:Deconv5}
{\rho }_{\mathrm{T}}\left(E\right)={\rho }_0\textrm{Re}\left[\frac{E-i\Gamma}{\sqrt{{\left(E-i\Gamma\right)}^2-{\Delta }^2_T}}\right], 
\end{equation}
where $\rho_{0}$ is the normal conducting DOS of the tip, and $\textrm{Re}$ indicates the real part.
We used the following parameters: \textit{T} = 0.3 K, for \textit{$\rho$}${}_{\textrm{T}}$(\textit{E}) we assume a superconducting DOS with a ${\Delta }_{\textrm{T}}=1.43$ meV, the $\Gamma$ parameter is set to $10^{-3}$~meV.

\subsection{Real-space simulation model}

For the real-space simulations, we use the following tight-binding Hamiltonian,
\begin{equation}\label{eq:TBHam}
\begin{split}
    \hat{H} =& \sum_{n=1}^2 \sum_{\langle \vec{r}, \vec{r}' \rangle_{n}}
    \Psi_{\vec{r}}^\dagger
    \left[
        t_{n} \sigma_0
        -i \alpha_{n} \left( \delta_y \sigma_x - \delta_x \sigma_y \right)        
    \right]\tau_{z}
    \Psi_{\vec{r}'}+ \mathrm{H.c.} \\
    +& \sum_{\vec{r}}
    \Psi_{\vec{r}}^\dagger
    \left[
        -\mu \sigma_{0}\tau_{z}
        + \Delta \sigma_{0}\tau_{x}
        - J \vec{S}_{\vec{r}} \cdot \vec{\sigma}\tau_{0}
    \right]
    \Psi_{\vec{r}},
\end{split}
\end{equation}
where $\langle \dots \rangle_{n}$ denotes summation over $n^{\text{th}}$-nearest neighbors,
\mbox{$\Psi_{\vec{r}}=\begin{pmatrix}\psi_{\vec{r}\uparrow},\psi_{\vec{r}\downarrow},\psi^{\dagger}_{\vec{r}\downarrow},-\psi^{\dagger}_{\vec{r}\uparrow}\end{pmatrix}^T$}, $t_{n}$ and $\alpha_{n}$ are hopping and Rashba parameters for the $n^\text{th}$-nearest neighbors, H.c. denotes Hermitian conjugate,
and the hopping direction is given by $\vec{\delta} = \left(\vec{r} - \vec{r}'\right)/\left|\vec{r} - \vec{r}'\right|$. $\tau$ and $\sigma$ matrices denote Pauli matrices, with the $0$ index standing for the identity matrix, in particle-hole and spin space, respectively. During the calculations, we consider the limit $J\gg t_{n},\alpha_{n},\mu+J,\Delta$, where two of the four bands in the model are located in the vicinity of the Fermi level, which is relevant for modeling the experimentally observed in-gap spectral features.
These two bands may be described by a Hamiltonian of the form in Eq.~\eqref{eq:GiannisEffHam}, with the choice of parameters $\epsilon_{\pm}\approx\pm\left(\mu+J\right)$, $s_{z,n}\approx-2t_{n}/J$, $s_{0,n}\approx-2\alpha_{n}/J$, $s_{y,n}\approx-4\alpha_{n}\Delta/J^{2}$.
We considered nearest and next-nearest neighbors in the model $n=1,2$ to obtain reasonable agreement with the experiments; see Appendix~\ref{sec:simulations} and Figs.~\ref{figS1} and \ref{figS2} for the influence of other parameter choices on the spectral features. 
Because of this approximation, the operators $\psi_{\vec{r}s}$ and the coupling constants in Eq.~\eqref{eq:TBHam} are meant do describe effective low-energy degrees of freedom, instead of microscopic degrees of freedom in the Fermi liquid such as spins and atomic orbitals, and their particle and hole parts have a different meaning.
The simulations are performed on a triangular lattice of magnetic sites with the lattice constant $l=2.48$~\AA, which is three-fourth of the Nb bulk lattice constant based on the structural model in Fig.~\ref{fig2}.
We chose an in-plane magnetization direction $\vec{S}_{\vec{r}}\parallel[001]$ following the magnetic ground state determined from first-principles calculations.  
We consider the island shape and size from Fig.~\ref{fig4} to calculate the spectrum of the Hamiltonian and the spatial distribution of the local density of states (LDOS).
We diagonalize the Hamiltonian to find the eigenvectors and the eigenvalues.

To calculate the LDOS, we take the absolute value squared of the eigenvectors, sum up over spins, and sum up over the energies of the eigenstates using a Lorentzian broadening of $\Delta E=0.002$~meV. This results in particle and hole density of states from the parts of the eigenvectors corresponding to annihilation $\psi_{\vec{r}s}$ and creation $\psi^{\dag}_{\vec{r}s}$ operators, respectively. We take the weighted sum of the particle and hole parts as: 
\begin{equation}
\textrm{LDOS}=P\cdot\textrm{LDOS}_{\textrm{particle}}+(1-P)\cdot\textrm{LDOS}_{\textrm{hole}},
\end{equation}
where $P$ is meant to represent the particle weight of the state described by the $\psi_{\vec{r}\uparrow}$ operator expressed in the microscopic fermionic degrees of freedom, while $1-P$ is the same type of particle weight for the state described by the $-\psi^{\dagger}_{\vec{r}\uparrow}$ operator.
This is similar in spirit to the procedure in Ref.~\cite{Schneider2021}, where the effective low-energy degrees of freedom represent single-impurity YSR states at positive and negative energies, which also have certain particle weights in the microscopic degrees of freedom.
Unfortunately, we have no information on the value of the $P$ parameter, since single Co adatoms on Nb(110) do not display clearly identifiable YSR states inside the gap~\cite{KusterNAT-COMM2021}, and the YSR bands are only formed as the Co superstructure is grown on the surface.
Therefore, we considered the value $P=0.5$, resulting in a symmetry between negative and positive energies in the simulations.

\backmatter

\bmhead{Acknowledgments}
R.L.C. acknowledges financial support from the Deutsche Forschungsgemeinschaft (DFG, German Research Foundation), Grant No. 459025680.
K.v.B. acknowledges funding from the Deutsche Forschungsgemeinschaft (DFG, German Research Foundation) under project number 552644472.
L.R. gratefully acknowledges funding by the National Research, Development, and Innovation Office (NRDIO) of Hungary under Project No. FK142601 and by the Hungarian Academy of Sciences via a J\'{a}nos Bolyai Research Grant (Grant No. BO/00178/23/11).
L.R. and L.S. acknowledge funding from NRDIO-Hungary ADVANCED Grant No. 149745.
K.P. acknowledges support from NRDIO-Hungary ADVANCED Grant No. 152473.
I.I., E.M., T.P., J.W. and R.W. acknowledge funding by the Cluster of Excellence ‘Advanced Imaging of Matter’ (EXC 2056, project ID 390715994) of the Deutsche Forschungsgemeinschaft (DFG).
T.P. acknowledge funding from the European Union (ERC Starting Grant QUANTWIST, project number 101039098).
R.W. acknowledges funding from the European Union through the ERC Advanced Grant ADMIRE (project number 786020).
J.W. acknowledges financial support from the Deutsche Forschungsgemeinschaft (DFG, German Research Foundation) via project WI 3097/4-1 (project number 543483081).
S.R. acknowledges support from the Australian Research Council through Grant No. DP240100168.

\bmhead{Author contributions}

R.L.C. and M.B. conceived the study. R.L.C., M.B., and P.B. performed the experiments and analyzed the data together with K.v.B. and J.W. R.W. supervised the low-temperature STM and STS investigations. K.P. performed the VASP calculations and the STM simulations in section 2.1. A.D. and L.S. performed the SKKR calculations for the magnetic interaction parameters in section 2.1; A.D., L.R. and L.S. determined the magnetic ground state based on these parameters. I.I. performed the calculations, topological analysis, investigation of type I and type II Weyl nodes and created the figures for the low-energy theory in section 2.2, supervised by and in discussion with T.P. Both contributed to the general discussion and the theory discussions. E.M. and L.R. performed the calculations for the real-space simulations in section 2.4. S.R. supervised the calculations for real-space simulations performed at the University of Melbourne. R.L.C., M.B., L.R., I.I., and E.M. wrote the manuscript. R.L.C. coordinated the writing of the manuscript. All authors commented on the manuscript.

\begin{appendices}

\section{First-principles calculations}\label{sec:DFT}

The geometry optimization was performed using the VASP package~\cite{Kresse1996,Kresse1996a,Hafner2008}.
The system was modelled by a $3\times 1$ supercell of the Nb(110) surface, with lattice constants of $\sqrt{2}a_{\textrm{Nb}}$ and $3a_{\textrm{Nb}}$ along the $[1\overline{1}0]$ and $[001]$ directions, respectively, where $a_{\textrm{Nb}}=3.3004$~\AA~is the lattice constant of bulk Nb.
The exchange-potential construction proposed by Perdew, Burke and Ernzerhof~\cite{Perdew1996} was used for the calculations, and a $7\times 3\times 1$ Monkhorst-Pack $\boldsymbol{k}$ mesh was used in reciprocal space.
The supercell contained 6 Nb atoms in each layer, while 8 Co atoms were considered in the Co layer.
The simulated slab consisted of four layers of Nb and a single layer of Co, with a vacuum of around $14$~\AA~thickness between the supercells along the direction perpendicular to the surface.
The atoms in the top Nb and in the Co layers were relaxed along all spatial directions, while the atoms in the second Nb layer were allowed to relax vertically. 
The energetically most favourable configuration is shown in Fig.~\ref{fig2}a.
Based on the atomic positions, the surface belongs to the \textit{pmg} symmetry group, reduced from the \textit{pmm} symmetry of the bcc(110) surface by the mirror plane with normal vector along $[1\overline{1}0]$ being reduced to a glide reflection, but preserving the twofold rotational symmetry around the out-of-plane direction relevant to the topological classification.
The corrugation in the geometry is below 1~pm in the relaxed Nb layers and below 5~pm in the Co layer, which was neglected in the following calculations.
The Co layer relaxed inwards by 11.3\% on average compared to the ideal Nb(110) interlayer distance, the top Nb layer relaxed outwards by 2\%, while the relaxation of the subsurface Nb layer is negligible.
The obtained atomic configuration had an average magnetic moment of 1.31~$\mu_{\textrm{B}}$ per Co atom with a difference of less than 10\% between the sites.

The above geometry was included in the screened Korringa--Kohn--Rostoker code~\cite{Szunyogh1995}.
The simulated surface consisted of nine layers of Nb, a single layer of Co and three layers of empty spheres (vacuum) between semi-infinite bulk Nb and semi-infinite vacuum.
The exchange-correlation potential was parametrized based on the method by Vosko, Wilk and Nusair~\cite{Vosko1980}.
256 $\boldsymbol{k}$ points were considered for the two-dimensional Brillouin zone integration due to the relatively large size of the atomic unit cell.
The energy integration was performed along a semicircle contour in the complex plane containing $12$ energy points.
An angular-momentum cutoff of $l_{\textrm{max}}=2$ was applied.
The spin magnetic moments of the Co atoms are found to be around 1.35~$\mu_{\textrm{B}}$, in reasonable agreement with the VASP calculations.
Tensorial exchange interactions containing Heisenberg, Dzyaloshinsky--Moriya and two-site anisotropy contributions in a classical spin model were calculated between pairs of Co atoms within a distance of $5a_{\textrm{Nb}}$ using ferromagnetic reference states along different crystallographic directions.
The obtained interaction coefficients were symmetrized assuming the \textit{pmg} symmetry to eliminate the numerical errors which could influence the in-plane orientation of the spins due to the weak in-plane anisotropy.

The magnetic ground state shown in Fig.~\ref{fig2}b was determined from atomistic spin dynamics simulations using the obtained spin model.
The system relaxed to this ground state even when initialized in a random initial configuration at zero temperature on a lattice containing $16\times 32$ atomic unit cells, indicating that the formation of non-collinear spin structures is energetically unfavorable.

\section{Atomic positions in the reconstructed structure}\label{sec:reconstruction}

\begin{figure}[ht!]
\centering
\includegraphics[width=1\linewidth]{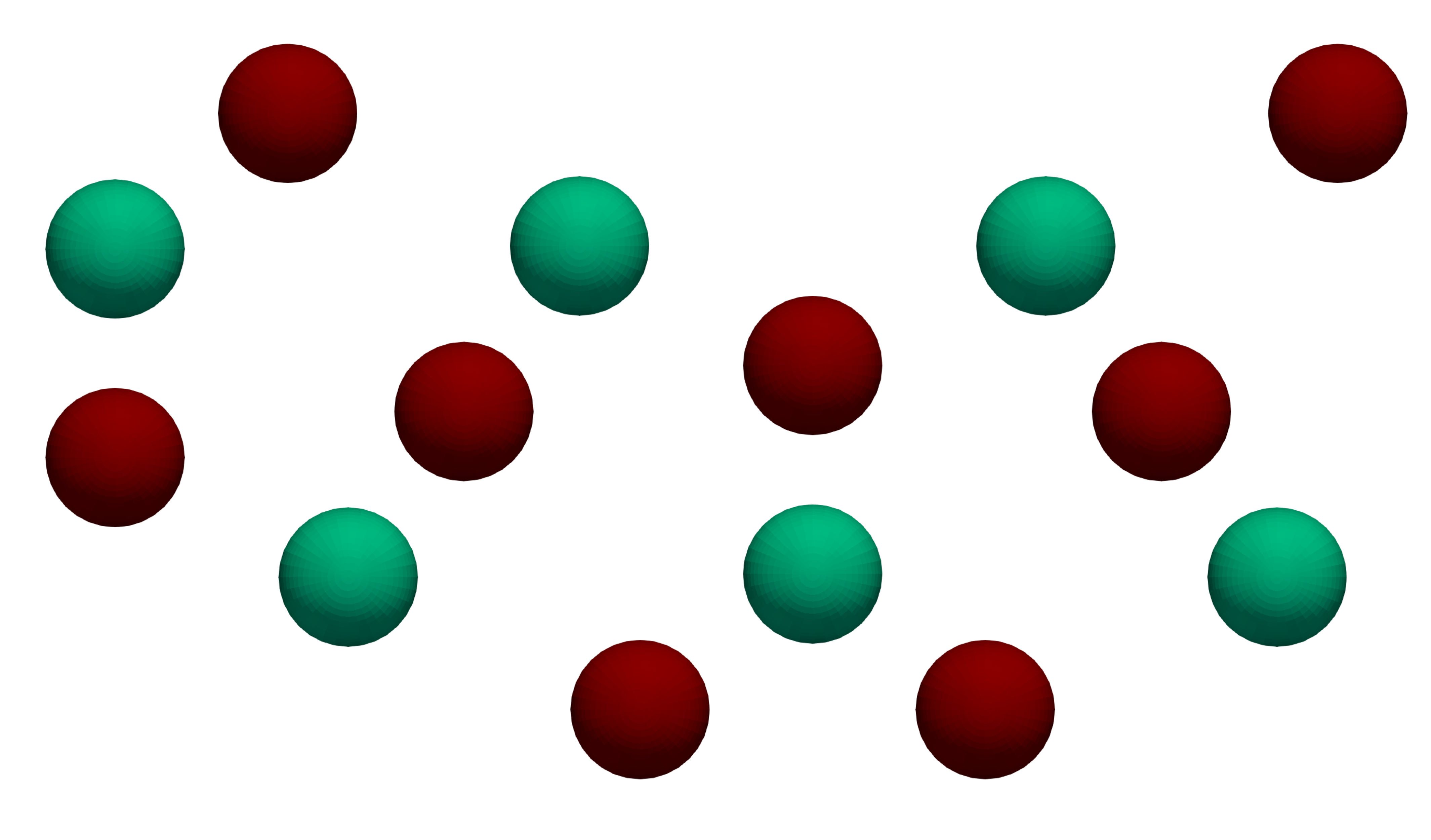}
\caption{
    Atomic positions inside the structural unit cell. Cyan and red spheres denote the Nb atoms in the top layer and the Co atoms, respectively.
}\label{figS3}
\end{figure}

The positions of the atoms in the top Nb layer and the Co layer in the relaxed reconstructed structure obtained from VASP calculations are reported in Table~\ref{tab:table1}.
The atoms shown in Fig.~\ref{fig2} in the main text are placed at these positions.
In the ideal bulk Nb geometry, the Nb atoms in this layer occupy the positions $\left(m/6,(2n+1)/4,0.3\right)$ for $m\in\left\{0,\dots,5\right\}$ and $n\in\left\{0,\dots,1\right\}$.
After the relaxation, the positions along the $x$ axis change by less than $0.1\%$ of the lattice constant of the supercell, along the $y$ axis slightly more but still by less than $0.2\%$ of the corresponding lattice constant, while along the $z$ axis they relax outwards by about $2\%$ of the equilibrium interlayer distance, which is one tenth of the supercell size along this direction.
The second and sixth Co atoms in the table occupy bridge positions between two Nb atoms, as shown in Fig.~\ref{figS3}.
These have the highest position along the $z$ axis, and are responsible for the bright spots forming a rectangular structure in the simulated STM image in Fig.~\ref{fig2}a.
The other six Co atoms form almost equilateral triangles around the third and sixth Nb atoms, and the fourth and eighth Co atoms are positioned approximately at the centers of triangles formed by Nb atoms, see Fig.~\ref{figS3}.
Overall, the pmm symmetry of the bcc(110) surface is changed to a pmg symmetry in the reconstructed structure via replacing a mirror plane by a glide plane.
The second and sixth Co atoms are located on out-of-plane twofold rotation axes, and the glide plane passes through them.
The mirror planes in the $yz$ plane pass through the fourth or the eighth Co atoms.

\begin{table}
\caption{Relaxed atomic positions in the top Nb layer and the Co layer. The positions are given in units of the lattice vectors of the supercell, $\left(a_{1},a_{2},a_{3}\right)=\left(3,\sqrt{2},5\sqrt{2}\right)a_{\textrm{Nb}}$, with $a_{\textrm{Nb}}=3.3004$~\AA~the lattice constant of bulk Nb.}
\label{tab:table1}
\begin{tabular}{rrrr}

      atom &          $x$ $(a_{1})$ &          $y$ $(a_{2})$ &          $z$ $(a_{3})$ \\
\hline
        Nb &     0.6671 &     0.7516 &     0.3021 \\

        Nb &     0.8329 &     0.2484 &     0.3021 \\

        Nb &     0.0000 &     0.7469 &     0.3016 \\

        Nb &     0.1671 &     0.2484 &     0.3021 \\

        Nb &     0.3329 &     0.7516 &     0.3021 \\

        Nb &     0.5000 &     0.2531 &     0.3016 \\

        Co &     0.6237 &     0.0470 &     0.3904 \\

        Co &     0.7500 &     0.5000 &     0.3920 \\

        Co &     0.8763 &     0.9530 &     0.3904 \\

        Co &     0.0000 &     0.4300 &     0.3898 \\

        Co &     0.1237 &     0.9530 &     0.3904 \\

        Co &     0.2500 &     0.5000 &     0.3920 \\

        Co &     0.3763 &     0.0470 &     0.3904 \\

        Co &     0.5000 &     0.5700 &     0.3898 \\

\end{tabular}  
\end{table}

\section{Details of the low-energy theory}\label{sec:Low-EnergyTheory}
In this section, we provide details for the derivation of the effective Hamiltonian in Eq. (\ref{eq:GiannisEffHam}) in the main text. Here, 
\begin{equation}\label{eq:GiannisDefinitions}
\begin{split}
k_{\textrm{F}}=& \sqrt{2 m \mu/\hbar^2},
\upsilon_{\textrm{F}}=\hbar k_{\textrm{F}}/m, \xi= \hbar \upsilon_{\textrm{F},\nu}/ \Delta, \rho_\pm=k^{\pm}_{\textrm{F}}/2\pi \upsilon_{\textrm{F}},\\
    k_{\textrm{F}\pm}=&k_{\textrm{F}} \left( \sqrt{1+\left(\dfrac{a}{\hbar \upsilon_{\textrm{F}}}\right)^2}\mp \dfrac{a}{\hbar \upsilon_{\textrm{F}}}\right),
    \upsilon_{\textrm{F},\nu}=\upsilon_{\textrm{F}} \left( \sqrt{1+\left(\dfrac{a}{\hbar \upsilon_{\textrm{F}}}\right)^2}\right),\\
        s_{z}(r)= &\dfrac{e^{-\frac{r}{\xi}}}{\sqrt{2\pi r}\upsilon_{\textrm{F}}} \left( \sqrt{k_{\textrm{F}}^+}\cos(k_{\textrm{F}}^+r-\pi/4)+ \sqrt{k_{\textrm{F}}^-}\cos(k_{\textrm{F}}^-r-\pi/4)\right),\\
         s_{y}(r)= &\dfrac{e^{-\frac{r}{\xi}}}{\sqrt{2\pi r}\upsilon_{\textrm{F}}} \left( \sqrt{k_{\textrm{F}}^+}\cos(k_{\textrm{F}}^+r-\pi/4)- \sqrt{k_{\textrm{F}}^-}\cos(k_{\textrm{F}}^-r-\pi/4)\right),\\
        \Tilde{s}(r)= &\dfrac{e^{-\frac{r}{\xi}}}{\sqrt{2\pi r}\upsilon_{\textrm{F}}} \left( \sqrt{k_{\textrm{F}}^+}\sin(k_{\textrm{F}}^+r-\pi/4)+ \sqrt{k_{\textrm{F}}^-}\sin(k_{\textrm{F}}^-r-\pi/4)\right), \\
            s_{0}(r)=&\dfrac{e^{-\frac{r}{\xi}}}{\sqrt{2\pi r}\upsilon_{\textrm{F}}} \left( \sqrt{k_{\textrm{F}}^+}\sin(k_{\textrm{F}}^+r-\pi/4)- \sqrt{k_{\textrm{F}}^-}\sin(k_{\textrm{F}}^-r-\pi/4)\right) .
\end{split}
\end{equation}
The functions $s_i(r)$ are related to bulk-mediated couplings between impurities at different sites. They decay exponentially for distances larger than the superconductor's coherence length $\xi$, while for smaller distances the elements decay as the square-root of the distance. Specifically, we follow Ref.~\cite{Li2016} and write the matrix Green's functions of the bulk in the BdG basis $\mathcal{G}=\ll\begin{pmatrix}\psi_{\uparrow},\psi_{\downarrow},\psi^{\dagger}_{\downarrow},-\psi^{\dagger}_{\uparrow}\end{pmatrix}^\dagger;\begin{pmatrix}\psi_{\uparrow},\psi_{\downarrow},\psi^{\dagger}_{\downarrow},-\psi^{\dagger}_{\uparrow}\end{pmatrix}\gg$, in Zubarev's notation~\cite{Zubarev}, 
\begin{equation}\label{eq:GiannisGF}
 \begin{split}
    \mathcal{G}(E,r)=&\tau_x \otimes w_x(r) + \tau_z \otimes w_z(r),\\ \mathcal{G}(E,0)=&\left(- \dfrac{E}{\Delta}\dfrac{\pi}{2}\sum_b \rho_b\right)\tau_0 \otimes \sigma_0 +\left(- \dfrac{\pi}{2}\sum_b \rho_b \right)\tau_x \otimes \sigma_0,\\
    w_x(r)=&-\dfrac{1}{2}s_z(r) \sigma_0-\dfrac{i}{2} s_0(r) \left( \sigma_y \cos(\phi_r) -\sigma_x \sin(\phi_r)\right),\\
     w_z(r)=& \dfrac{1}{2}s_0(r) \sigma_0-\dfrac{i}{2} s_z(r) \left( \sigma_y \cos(\phi_r) -\sigma_x \sin(\phi_r)\right).
   \end{split}    
 \end{equation}
 We aim to solve the eigenvalue equation
 \begin{equation}\label{eq:GiannisEigenEquation}
    \sum_{r'} \mathcal{G}\left(E,r-r'\right) \left(-J \tau_0 \otimes \vec{S}\vec{\sigma} \right)\psi(r')=\psi(r),
\end{equation}
where $\Vec{S}=S \begin{pmatrix}  \cos(\theta)\sin(\zeta),\sin(\theta)\sin(\zeta),\cos(\zeta) \end{pmatrix}$. After projecting the wave function to the bare YSR wave functions $
    \psi_+= \begin{pmatrix}
    x_{\uparrow}, &
    x_{\uparrow}
    \end{pmatrix}^T, \psi_-=\begin{pmatrix}
    x_{\downarrow}, &
    -x_{\downarrow}
    \end{pmatrix}^T$, where we define $x_{\uparrow}=\begin{pmatrix}
        \cos(\zeta/2)e^{-i\frac{\theta}{2}}, &
         \sin(\zeta/2)e^{i\frac{\theta}{2}}
    \end{pmatrix}^T/\sqrt{2}, x_{\downarrow}=\begin{pmatrix}
         \sin(\zeta/2)e^{-i\frac{\theta}{2}}, &
         -\cos(\zeta/2)e^{i\frac{\theta}{2}}
    \end{pmatrix}^T /\sqrt{2}$, taking the limit $E/\Delta \rightarrow 0$, and transforming to Fourier space, we reach at Eq.~(\ref{eq:GiannisEffHam}).

To examine the robustness of the Weyl cone formation and provide analytical insights into the type of Weyl cones that form in the Brillouin zone, we simplify our model by considering a nearest-neighbor real-space cut-off. 
Similar studies with truncated YSR hopping integrals have revealed good agreement with experimental results~\cite{Schneider2021}.
In this case, the matrix elements in the general form in Eq.~(\ref{eq:GiannisEffHam}) simplify to

\begin{equation}\label{eq:2DYSRVectorComponentsSimplified}
    \begin{split}
        d_z(\Vec{q})=&J s_z(l) \left(\cos(l q_x)+ \cos\left(\dfrac{l}{2}(q_x+\sqrt{3}q_y)\right)+ \cos\left(\dfrac{l}{2}(-q_x+\sqrt{3}q_y)\right) \right), \\
         d_y(\Vec{q})=&-Js_y(l) \left(\cos(\theta) \sin(l q_x) + \cos\left(\theta-\dfrac{\pi}{3}\right)\sin\left(\dfrac{l}{2}(q_x+\sqrt{3}q_y)\right)\right.\\&\left.- \cos\left(\theta+\dfrac{\pi}{3}\right)\sin\left(\dfrac{l}{2}(-q_x+\sqrt{3}q_y)\right)\right),\\  
         d_0(\Vec{q})=&-Js_0(l) \left(\sin(\theta) \sin(l q_x) + \sin\left(\theta-\dfrac{\pi}{3}\right)\sin\left(\dfrac{l}{2}(q_x+\sqrt{3}q_y)\right)\right.\\&\left.- \sin\left(\theta+\dfrac{\pi}{3}\right)\sin\left(\dfrac{l}{2}(-q_x+\sqrt{3}q_y)\right) \right),
    \end{split}
\end{equation}
where we explicitly consider the triangular lattice primitive vectors $\Vec{r}_1= l \begin{pmatrix}
        1, & 0 
    \end{pmatrix},$ and $ \Vec{r}_2= l \begin{pmatrix}
        1, & \sqrt{3} 
    \end{pmatrix}/2$.
In the case where the magnetization direction aligns with the crystallographic direction $\Vec{r}_1$, i.e., $\theta=0$, the gap closings of Eq.~(\ref{eq:2DYSRVectorComponentsSimplified}) are found at momenta $\Vec{q}_0=\pm(0,q_{y,0})$, where $q_{y,0}=4\pi \sqrt{3}/(9l)$.
To examine the spectrum next to the gap-closings, we linearize $\Vec{d}$ around $+\Vec{q}_0$ and get

\begin{equation}\label{eq:dvectorlinearized}
\begin{split}
    &d_0(\delta \Vec{q})=d_0(0)+v_{y,0}(q_y-q_{y,0}), d_y(\delta \Vec{q})=v_{x,y}q_x, d_z(\delta \Vec{q})=v_{y,z}(q_y-q_{y,0}),\\
    & d_0(0)=\dfrac{3Js_0}{2},v_{y,0}=\dfrac{3Js_0l}{4}, v_{x,y}=-\dfrac{5Js_yl}{4},  v_{y,z}=-\dfrac{3Js_zl}{2},
    \end{split}
\end{equation}
and, consequently, the low-energy excitation spectrum near the Weyl cone,
\begin{equation}\label{eq:dispersionlinearized}
    E_{\pm}(\Vec{q})=d_0(0)+v_{y,0}(q_y-q_{y,0})\pm \sqrt{\left(v_{x,y}q_x\right)^2+\left(v_{y,z}(q_y-q_{y,0})\right)^2}.
\end{equation}
In the above, $d_0$ lifts the Weyl cones from zero energy and the term $v_{y,0}(q_y-q_{y,0})$ tilts the Weyl cones in the direction that connects the pair of cones, see Fig.~\ref{fig5}.
In the $y$ direction, the Weyl cones can then be classified to type I or type II depending on the relative strength, $\mathcal{S}={\rm sign}(|v_{y,z}|-|v_{y,0}|)$.
A type II Weyl cone is, therefore, formed when $\mathcal{S}<0$, or equivalently, $|2s_z(l)|<|s_0(l)|$. From Eq.~(\ref{eq:GiannisDefinitions}) we note that the above condition sensitively depends on the microscopic parameters of the system and in particular the lattice constant of the triangular lattice $l$.
    
\section{Simulations for different parameter choices}\label{sec:simulations}

We conducted calculations with the real-space tight-binding model to investigate the influence of the strength of the spin--orbit coupling on the calculated LDOS.
Assuming that the magnetic moments are pointing along the $x$ direction at every lattice site, Eq.~\eqref{eq:TBHam} in the Methods may be expressed in Fourier space as 
\begin{align}
    \hat{H} =& \sum_{\vec{q}}
    \Psi_{\vec{q}}^\dagger
    \left[
        -\mu+t\left(\vec{q}\right) \sigma_{0}\tau_{z}
        -i \alpha_{y}\left(\vec{q}\right)\sigma_{x}\tau_{z}
        \right.
        \nonumber
        \\
        &\left.
        +i \alpha_{x}\left(\vec{q}\right)\sigma_{y}\tau_{z} +\Delta\sigma_{0}\tau_{x}+J\sigma_{x}\tau_{0}  
    \right]
    \Psi_{\vec{q}},\label{eq:FourierHam}
\end{align}
where $t\left(\vec{q}\right), i \alpha_{y}\left(\vec{q}\right)$ and $i \alpha_{x}\left(\vec{q}\right)$ denote the Fourier transforms of the hopping terms, as well as the Rashba terms along the $y$ and $x$ spatial directions.
For the nearest-neighbor terms, these have the same functional dependence on the wave vector as $d_{z}\left(\vec{q}\right),d_{0}\left(\vec{q}\right)$ and $d_{y}\left(\vec{q}\right)$ in Eq.~\eqref{eq:2DYSRVectorComponentsSimplified}, respectively.
We introduce the notations
\begin{align}
\delta E_{0}\left(\vec{q}\right)=&\left(J-\sqrt{\Delta^{2}+\left(-\mu+t\left(\vec{q}\right)-i \alpha_{y}\left(\vec{q}\right)\right)^{2}}\right)
\nonumber \\
&-\left(\sqrt{\Delta^{2}+\left(-\mu+t\left(\vec{q}\right)+i \alpha_{y}\left(\vec{q}\right)\right)^{2}}-J\right),\label{eq:deltaE0}\\
\tilde{d}_{y}\left(\vec{q}\right)=&i \alpha_{x}\left(\vec{q}\right).\label{eq:tildedy}
\end{align}
along the contours where $\tilde{d}_{y}\left(\vec{q}\right)=0$, $\delta E_{0}\left(\vec{q}\right)$ is equal to the energy difference between the two eigenstates closest to the Fermi level in the limit of $J\gg t_{n},\alpha_{n},\mu+J,\Delta$.
Note that these two states are not symmetric to the Fermi level at a given $\vec{q}$ point due to the sign change in the $i \alpha_{y}\left(\vec{q}\right)$, similarly to the tilting induced by the $d_{0}\left(\vec{q}\right)$ term in Eq.~\eqref{eq:GiannisEffHam}.
The crossing points of the nodal lines of $\delta E_{0}\left(\vec{q}\right)$ and $\tilde{d}_{y}\left(\vec{q}\right)$ give the positions of the Weyl nodes where the two energy levels closest to the Fermi energy are degenerate, following the analogy with Eq.~\eqref{eq:GiannisEffHam} where $d_{z}\left(\vec{q}\right)$ and $d_{y}\left(\vec{q}\right)$ played the same role.

\begin{figure}[ht!]
\centering
\includegraphics[width=1\linewidth]{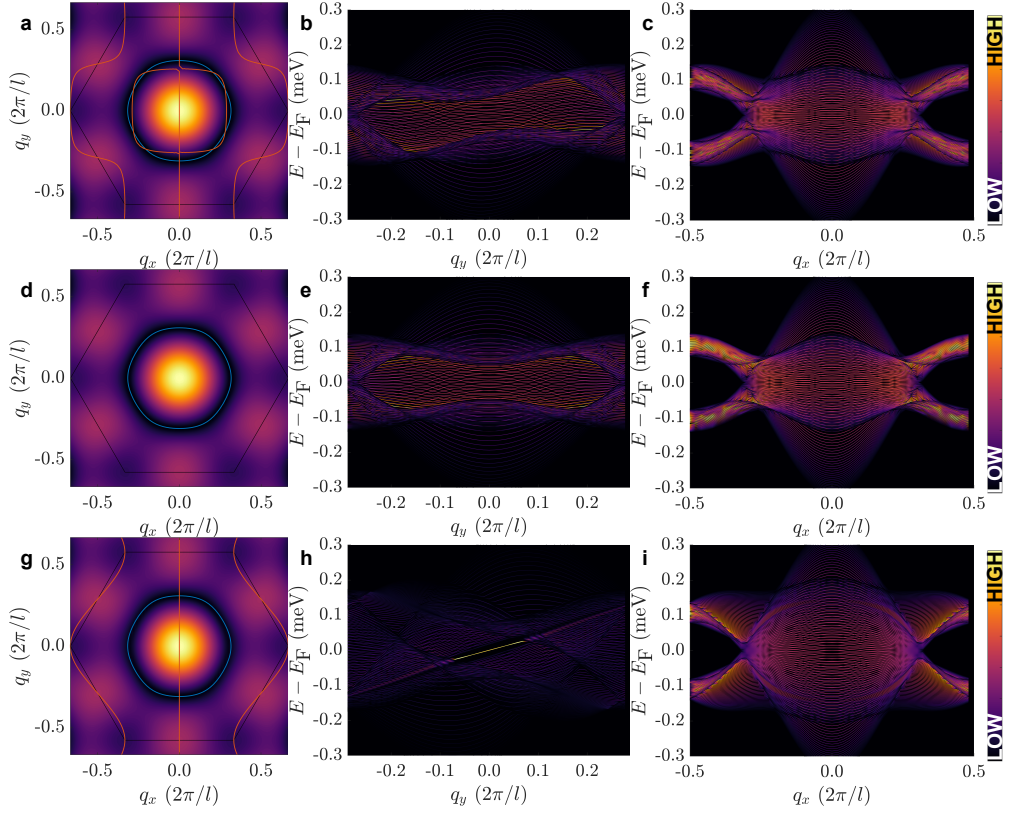}
\caption{
    Calculation results of the tight-binding model in reciprocal space and in slab geometries.
    \textbf{a},\textbf{d} and \textbf{g} Contour plot of the energy splitting $\delta E$ of the two bands in the vicinity of the Fermi level from Eq.~\eqref{eq:FourierHam}, similarly to Fig.~\ref{fig5}c in the main text. The black hexagon represents the Brillouin zone. The blue and orange lines are nodal lines of $\delta E_{0}\left(\boldsymbol{q}\right)$ and $\tilde{d}_{y}\left(\boldsymbol{q}\right)$ from Eqs.~\eqref{eq:deltaE0} and \eqref{eq:tildedy}, respectively. \textbf{b},\textbf{e} and \textbf{h} Calculated spectra for slab geometries: 
    periodic boundary conditions are applied along the $y$ direction, and the slab is $50$ sites wide along the $x$ direction. Color denotes the weight of a given state on the edge atom of the slab. \textbf{c},\textbf{f} and \textbf{i} Calculated spectra for slab geometries: periodic boundary conditions are applied along the $x$ direction, and the slab is $50$ sites wide along the $y$ direction. 
    Parameters used for the simulations in Eq.~\eqref{eq:TBHam} are $t_{1}=41.7~\mu$eV, $t_{2}=16.7~\mu$eV, $\mu+J=30.0~\mu$eV, $\Delta=100~\mu$eV, and $J=1000~\mu$eV. 
    The spin--orbit coupling was varied as \textbf{a}-\textbf{c} $\alpha_{1}=-3.33~\mu$eV, $\alpha_{2}=8.33~\mu$eV; 
    \textbf{d}-\textbf{f} $\alpha_{1}=0.00~\mu$eV, $\alpha_{2}=0.00~\mu$eV; 
    and \textbf{g}-\textbf{i} $\alpha_{1}=41.7~\mu$eV, $\alpha_{2}=0.00~\mu$eV. 
}\label{figS1}
\end{figure}

The spectrum of Eq.~\eqref{eq:FourierHam} is illustrated in Fig.~\ref{figS1} for different choices of the spin--orbit coupling parameters.
The energy difference $\delta E$ between the two bands closest to the Fermi level is reduced along a circle around the $\Gamma$ point for the given parameter choice.
This circle closely aligns with the line $\delta E_{0}\left(\boldsymbol{q}\right)=0$ denoted by blue color, and coincides with it in the absence of spin--orbit coupling in Fig.~\ref{figS1}d, forming a nodal line.
This nodal line is reduced to a set of Weyl nodes if $\tilde{d}_{y}\left(\boldsymbol{q}\right)$ is finite.
The shape of the orange lines $\tilde{d}_{y}\left(\boldsymbol{q}\right)=0$, therefore the number of Weyl points where they intersect the contour $\delta E_{0}\left(\boldsymbol{q}\right)=0$, depends on the choice of parameters: there are two Weyl points for only nearest-neighbor Rashba terms in Fig.~\ref{figS1}g, but there are ten Weyl points when next-nearest-neighbor Rashba terms are taken into account in Fig.~\ref{figS1}a, using the same parameters as in the main text.

\begin{figure}[ht!]
\centering
\includegraphics[width=1\linewidth]{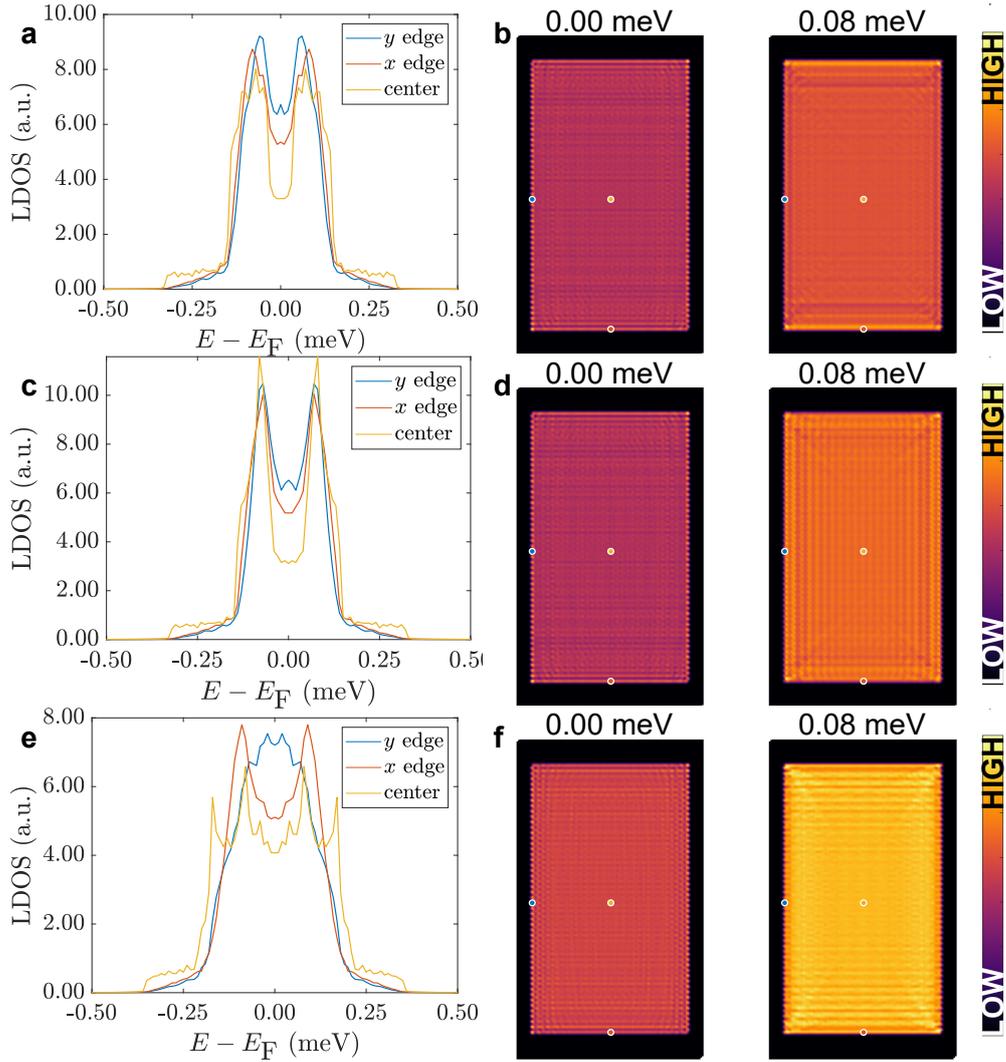}
\caption{
    Calculated spectrum of a rectangular island with different choices of parameters.
    \textbf{a},\textbf{c} and \textbf{e} Calculated LDOS at the edge along the $y$ direction (blue), at the edge along the $x$ direction (orange) and at the center (yellow) of the island. The positions of the atoms are shown as correspondingly colored dots in panels \textbf{b},\textbf{d} and \textbf{f}.
    \textbf{b},\textbf{d} and \textbf{f} Two-dimensional LDOS map at the energies $E-E_{\textrm{F}}$ indicated above the panels.
    Parameters used for the simulations in Eq.~\eqref{eq:TBHam} are $t_{1}=41.7~\mu$eV, $t_{2}=16.7~\mu$eV, $\mu+J=30.0~\mu$eV, $\Delta=100~\mu$eV, and $J=1000~\mu$eV. 
    The spin--orbit coupling was varied as \textbf{a}-\textbf{c} $\alpha_{1}=-3.33~\mu$eV, $\alpha_{2}=8.33~\mu$eV; 
    \textbf{d}-\textbf{f} $\alpha_{1}=0.00~\mu$eV, $\alpha_{2}=0.00~\mu$eV; 
    and \textbf{g}-\textbf{i} $\alpha_{1}=41.7~\mu$eV, $\alpha_{2}=0.00~\mu$eV. 
}\label{figS2}
\end{figure}

The role of the different terms may be further discussed based on the spectrum of a finite-width slab, shown in Fig.~\ref{figS1}.
In the absence of Rashba terms in Fig.~\ref{figS1}e and f, the spectrum is symmetric in wave vector in slabs with infinite edges running along the $y$ and $x$ directions, and no relative gap is visible close to the center of the Brillouin zone due to the nodal line.
The panels are colored based on the weight of the wave functions on the edge atom, showing a general enhancement between $-0.08$~meV and $0.08$~meV relative to the Fermi level.
In the presence of Rashba terms, the spectrum becomes tilted for the slab with edge along the $y$ direction in Fig.~\ref{figS1}b, but remains symmetric for the edge running along the $x$ direction in Fig.~\ref{figS1}c.
This is a consequence of the magnetic moments lying along the $x$ direction, which cannot cause a tilting in geometries with edges parallel to the magnetization direction.
Still no relative gap is visible at low wave vectors due to the interplay between the tilting, the high number of Weyl points, and the low value of the spin--orbit coupling.
Increasing the spin--orbit coupling enhances the tilting in Fig.~\ref{figS1}h, while the spectrum remains symmetric for the other edge orientation in Fig.~\ref{figS1}i as mentioned above.
Furthermore, a small relative gap opens for $q_{y}\in\left[-0.08,0.08\right]2\pi/l$, and edge states with a very high intensity are visible inside this gap in Fig.~\ref{figS1}h.
These edge states correspond to the Fermi arc connecting the tilted Weyl nodes.
Due to the tilting of the Weyl nodes, the Fermi arc is not completely flat in energy, unlike in the absence of spin--orbit coupling~\cite{Godzik2020}.
Such edge states are not visible in Fig.~\ref{figS1}i where both Weyl nodes are located at $q_{x}=0$.

The shape of the dispersion and the presence or absence of edge states influences the LDOS measured on finite islands, as illustrated in Fig.~\ref{figS2}.
Even in the absence of Rashba terms where a nodal line is formed, the intensity at the Fermi level is enhanced along edges both along the $x$ and $y$ directions, as shown in Fig.~\ref{figS2}c and d.
However, the intensity in the peaks at $E-E_{\textrm{F}}=\pm 0.08$~meV is slightly lower at the edges than inside the island, which is also visible in the two-dimensional map of the LDOS in Fig.~\ref{figS2}d.
Using the same parameters as in the main text in Fig.~\ref{figS2}a and b, the intensity along the edges is now enhanced in the whole range between $E-E_{\textrm{F}}=-0.08$~meV and $E-E_{\textrm{F}}=0.08$~meV, providing the best agreement with the experimental data.
This can be most likely attributed to the tilting, which broadens the peaks and also extends the enhancement along the edge to a wider energy range, as is also visible in Fig.~\ref{figS1}b and c compared to e and f.
Restricting the Rashba terms to nearest neighbors and further increasing the spin--orbit coupling leads to a large asymmetry along the different edges in Fig.~\ref{figS2}e and f: the LDOS is peaked at zero energy for the edge along the $y$ direction, while the two-peak structure is preserved for the edge along the $x$ direction.
This is a consequence of the emergence of edge modes along the $y$ direction, discussed above in the context of Fig.~\ref{figS1}h.
The peaks at the center of the island also split into pairs due to the enhanced tilting, and the LDOS takes similar values both inside the island and along the edges at $E-E_{\textrm{F}}=\pm 0.08$~meV, as shown in Fig.~\ref{figS2}f.
Since the single peak at the Fermi level was not observed along any of the edges in the experiments, clear edge states in the actual system are most likely suppressed, which can be reproduced by the simulations with a lower spin--orbit coupling and the combination of nearest-neighbor and next-nearest-neighbor Rashba terms resulting in more Weyl nodes, as shown in Fig.~\ref{figS1}a.

\end{appendices}

\bibliography{main}

\end{document}